\documentclass{aa}
\usepackage{graphicx}
\usepackage{txfonts}
\usepackage{xcolor}
\usepackage[]{changes}
\definechangesauthor[name=Dominique,color=orange]{dom}
\definechangesauthor[name=Julien]{jul}

\newcommand{\cmfast}{{\texttt{21cmFAST} }}

\newcommand{\ZT}{z_{\mathrm{obs}}}
\newcommand{\ZR}{z_{reion}(\Vec{r})}
\newcommand{\TR}{t_{\mathrm{reion}}(\Vec{r})}

\usepackage{enumitem}
\newcommand{\Cite}[1]{\textcolor{blue}{\cite{#1}}}

\usepackage{multirow}

\begin{document}

   \title{Reionisation time field reconstruction from 21 cm signal maps}

   \author{Julien Hiegel
          %\inst{1}
          \and
          Emilie Thélie
          \and
          Dominique Aubert%\inst{1}  
          \and
          Jonathan Chardin 
          \and
          Nicolas Gillet
          \and
          Pierre Galois
           \and
          Nicolas Mai
          \and
          Pierre Ocvirk
          \and
          Rodrigo Ibata
          }

   \institute{Université de Strasbourg, CNRS UMR 7550, Observatoire Astronomique de Strasbourg, Strasbourg, France\\
              \email{julien.hiegel@astro.unistra.fr}}

   \date{Received ..., accepted ...}

\abstract
   % Context
   {
   %\LEt{***General notes. a) You show a preference for UK English language conventions, and I have edited accordingly throughout. b) A\&A uses the past tense to describe the specific steps used in a paper and the present tense to describe general methods and recent findings. Please make sure my edits are accurate in this respect throughout the paper (see Sect. 6 of the Language Guide https://www.aanda.org/for-authors/language-editing/6-verb-tenses). c) All abbreviations and acronyms must be introduced at first use, once in the Abstract and again in the main text. Instrument and programme names must be introduced (when appropriate) in the main text. All abbreviations should be used consistently. Please check throughout. }
   During the epoch of reionisation, the intergalactic medium is reionised by the UV radiation from the first generation of stars and galaxies. One tracer of the process is the 21 cm line of hydrogen that will be observed by the Square Kilometre Array (SKA) at low frequencies, thus imaging the distribution of ionised and neutral regions and their evolution. }
    % Aims
   {To prepare for these upcoming observations, we investigate a deep learning method to predict from 21 cm maps the reionisation time field ($\TR$), the time at which each location has been reionised. The  $\TR$ method
   %\LEt{***algorithm? (to avoid starting the sentence with a single letter) } 
   encodes the propagation of ionisation fronts in a single field, and  gives access to times of local reionisation or to the extent of the radiative reach of early sources. Moreover it gives access to the time evolution of ionisation on the plane of sky, when this evolution is usually probed along the line-of-sight direction. }
   % Methods
   {We trained a convolutional neural network (CNN) using simulated 21 cm maps and reionisation time fields produced by the simulation code \cmfast. We also investigated the performance of the CNN when adding instrumental effects.}
   % Results
   {Overall,
   %\LEt{***"globally"suggests the globe=the Earth } 
   we find that without instrumental effects the 21 cm maps can be used to reconstruct the associated reionisation times field in a satisfying manner. The quality of the reconstruction is dependent on the redshift at which the 21 cm observation is being made, and in general it is found that small-scale features (<10cMpch$^{-1}$)  are smoothed in the reconstructed field, while larger-scale features are recovered well. When instrumental effects are included, the scale dependence of reconstruction is even further pronounced, with significant smoothing on small and intermediate scales.}
   %Conclusion
   {The reionisation time field can be reconstructed, at least partially, from 21 cm maps of IGM during the epoch of reionisation. This quantity can thus be derived in principle from observations, and should then provide a means
   %\LEt{***a means=a way; a mean=an average } 
   to investigate the effect of local histories of reionisation on the first structures that appear in a given region.}


   \maketitle
%
%--------------------------------------------------------------------
%--------------------------------------------------------------------
%--------------------------------------------------------------------
\section{Introduction}

One of the most important transitions in the history of the Universe is the epoch of reionisation (EoR), a period driven by collapsed dark matter halos where the first galaxies and stars emerge (\Cite{Loeb_2001}, \Cite{Wise2019}, \Cite{Dayal2018}, \Cite{JulianB2020}). The light emitted by these sources started to reionise the intergalactic medium (IGM), mainly composed of hydrogen. This phenomenon is often pictured as a network of growing ionised bubbles, where the centre of the  bubbles host the sources of light (\Cite{Furlanetto_2004}, \Cite{2022A&A...658A.139T}). Eventually, these growing regions percolate until the whole IGM gets reionised, ending the EoR near z=5.5-6 (e.g. \Cite{Kulkarni2019}, \Cite{Konno2014}).
%{This period is unfortunately still misunderstood due to a lack of direct observations. However, thanks to the tremendous amount of neutral hydrogen omnipresent in the universe, } 

This  epoch can be probed using the 21 cm signal produced by a spin-flip transition (\Cite{Furlanetto2006}). This process releases a photon with an initial frequency $f_0$ = 1420 MHz that will be redshifted until it reaches us. Such low-frequency radio observations allow us to infer EoR properties, for example from   the 21 cm power spectrum (e.g. \Cite{Furlanetto2004}, \Cite{Zaldarriaga2004}, \Cite{Mesinger2013}, \Cite{Iliev2012}, \Cite{Greig2017}, \Cite{Zhao2022}, \Cite{Nasirudin_2020}, \Cite{Pagano_2020}, \Cite{Gazagnes_2021}, \Cite{Liu_2016}, \Cite{Gorce_2023}) or the 21 cm bispectrum. (\Cite{Karagiannis_2022}, \Cite{hutter_2020}) For example, the Low Frequency Array\footnote{https://www.astron.nl/telescopes/lofar/} (LOFAR, \Cite{VanHaarlem}) sets upper limits on the 21 cm signal power spectrum, putting the first constraints on the state of the IGM  on the high emissivity of UV photons (\Cite{Ghara2020} or on the radio background (\Cite{Mondal_2020}).
%according to several scenarios thanks to low frequency observations with which upper limits  have been obtained. Such limits enlighten our understanding of the early universe and help us to constraint parameters that rules the EoR. For example high emissivity of UV photons would induce HII bubbles to be few but large (\Cite{Ghara2020}).
%or on the radio background of the universe (\Cite{Mondal_2020}). 
Likewise, the Hydrogen Epoch of Reionisation Array\footnote{http://reionisation.org/} (HERA)  is designed to study the 21 cm power spectrum to constrain several parameters such as the EoR timing  (\Cite{DeBoer_2017});  for example, it was recently able to put actual boundaries on the X-ray heating produced by the first galaxies (\Cite{Hera2022}).

The Square Kilometer Array\footnote{https://skatelescope.org} (SKA, see e.g. \Cite{mellema2013}), will soon be built, and will have  enough sensitivity, resolution, and coverage at low frequencies to measure the 21 cm signal at high redshift and map the hydrogen distribution during the EoR. While SKA will also be able to investigate the EoR from the 21 cm power spectrum, SKA will   give us the unique opportunity to get images of the HI state. Such observations at different frequencies, hence different redshifts, will not only track the HI in 2D on the sky, but also along the line of sight, providing the time evolution of the signal.

This tomography is a great opportunity to explore the EoR (e.g. \Cite{GiriThesis}, \Cite{mellema2015}). SKA will allow us  to study astrophysical parameters providing information on the IGM, size, and distribution of ionised bubbles or the properties of the first generation of galaxies  (e.g. \Cite{mellema2013}). %It will also help to look at the metallicity dependence of high X-ray binaries (e.g. \Cite{Kaur_2022}). 
By extension, 21 cm observations from the EoR would help to improve our understanding of the early universe and to constrain many of its facets, such as the optical depth $\tau$ of the last scattering surface (e.g. \Cite{Billings}) or the properties of sources and propagation of ionising photons (e.g. \Cite{Shaw}); these observations would also help us understand the properties of dark matter by studying the non-linear matter power spectrum (e.g. \Cite{Markus})
%.\LEt{***have I\ grouped correctly? (I had to avoid a syntax problem to maintain "parallel structure")}

In this spirit, the aim of this  paper is to investigate how these future 21 cm observations can help us to study how the reionising radiation propagated, and how it started and evolved. We    focus on finding the seeds of the ionising photons that set off the reionisation and on monitoring the propagation and eventual percolation of reionisation fronts.
%Such information will allow knowing when and where the ionised bubbles emerged and percolated.
\begin{figure*}[h!]
    \centering
    \includegraphics[width=1\textwidth]{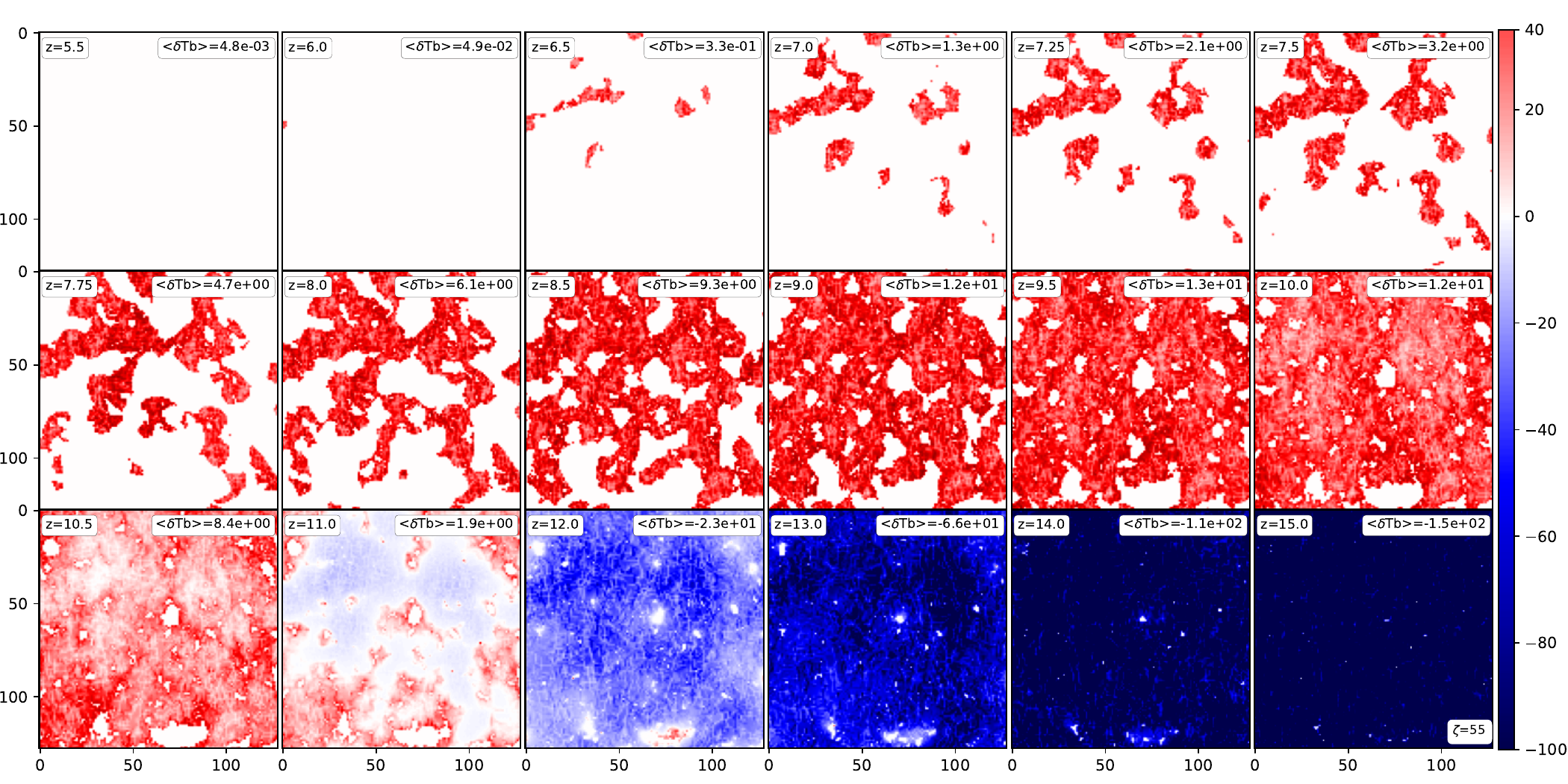}
    \caption{Timeline of $\delta T_b$. The colour bars are in units of [mK]; <$\delta T_b$> represents the mean of the temperature brightness. These maps come from a \cmfast simulation, taken at a given depth from the model $\zeta55$ and for each redshift used in this study. White corresponds to the absence of signal, meaning that the hydrogen is ionised at this region. For this simulation, at z=5.5, there is no neutral hydrogen anymore. Starting at z=15, $\delta T_b$ is mainly seen in absorption (negative values in blue) until z=11, where <$\delta T_b$> becomes positive and the signal is seen in emission (positive values in red). During the whole process, HII bubbles grow with time. Figure \ref{fig:zreionMAP} corresponds to the $\TR$ associated to these maps.}
    \label{fig:TB}
\end{figure*}

\begin{figure}
    \centering
    \includegraphics[width=0.5\textwidth]{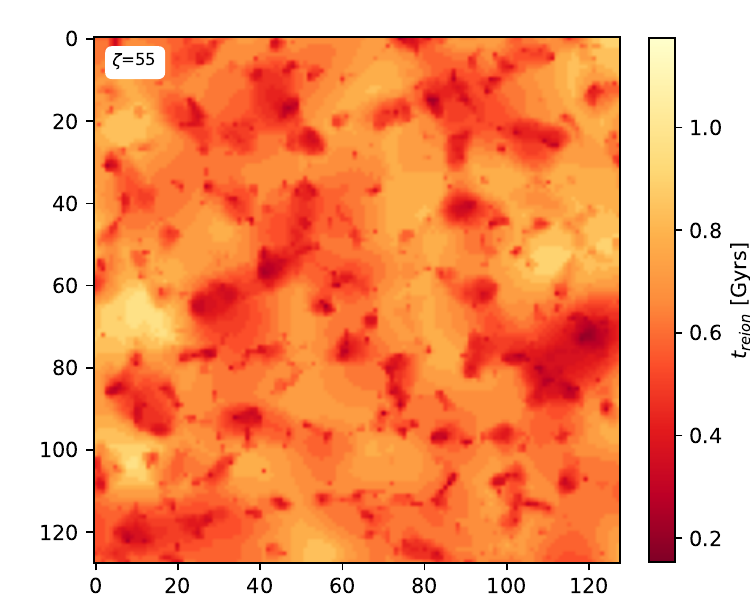}
    \caption{Example of 2D $\TR$ map from a $\zeta$=55 \cmfast model. The darker the region, the sooner it reionised. In this scenario (and for the whole $\zeta$55 dataset), the time of reionisation of the first HII regions is approximately 0.15 Gyr
    %\LEt{***a unit abbreviation is NEVER plural: five  years, 5 yr. Check in the figures, e.g. x-axes in Fig. 4    } 
    (z$\approx$20, darkest spots), and the last HI regions are reionised at around 1.1 Gyr (z$\approx$5.5, brightest regions). The mean value is 0.61 Gyr (z$\approx$8.4). This $\TR$ map is associated with Fig. \ref{fig:TB}, taken at the same depth in the simulation box.}
    \label{fig:zreionMAP}
\end{figure}

The 21 cm signal contains a significant amount of physical information, encoded by the temperature brightness $\delta T_b$ (see \Cite{Bianco2021}, \Cite{Prelogovic2021}, \Cite{Furlanetto2006}, \Cite{mellema2006}):
 \begin{multline}
\delta T_b(z) \approx  27 x_{HI}(z)(1+\delta_b (z))\left(\frac{1+z}{10}\right)^{\frac{1}{2}}
    \left(1 - \frac{T_{{\mathrm{CMB}}}(z)}{T_s(z)}  \right) \\
    \shoveleft{\left(\frac{\Omega_b}{0.044} \frac{h}{0.7}   \right)\left(\frac{\Omega_m}{0.27}  \right)^{-\frac{1}{2}}\text{[mK]}},
\end{multline}\\
which depends on the neutral fraction of hydrogen $x_{HI}$, the
density contrast of baryons $\delta_b$, the cosmic microwave background) 
%\LEt{***used only here so spell out CMB (see Note 1c) }
temperature $T_{\mathrm{CMB}}$, and the so-called spin temperature $T_s$ driven by the thermal state of the gas or the local amount of Ly-$\alpha$ radiation (\Cite{Liszt2001}). 

%\ddom{More the UV photons available more the coupling between these temperature, also true in high density regions (see \cite{mellema2013}). $T_s$ tells about the ratio of parallel to antiparallel spins, i.e the thermal state of the HI gas. $\delta T_b$ is directly proportional to the fraction of neutral hydrogen $x_{HI}$, that decreases with cosmological time until the gas is fully reionised. In addition, $\delta T_b$ depends on $\delta_b$, the density contrast of baryons. It is then possible to recover main sources of reionisation and the underlying structures specific to this epoch. More importantly, the whole history of reionisation is encoded in $\delta T_b$ via the redshift dependency of these quantities.}

A single 21 cm observation can therefore provide  direct insight into the state of these quantities at the observed redshift $z$. Figure \ref{fig:TB} shows examples of mock 21 cm observations, obtained thanks to \cmfast ((\Cite{Mesinger2011}, \Cite{Murray2020}), see Sect. \ref{sec:NeuralNetwork}).
%\ddom{ for each redshift used in this study}.
From z=15 to z=5.5 we can observe HII bubbles (in white), inside of which no signal can be observed, growing with time until only HII remains and the radio signal vanishes. Since each observation in this sequence is a snapshot of a propagation process, they are correlated. At the extreme, it can even be envisioned that a single 21 cm observation may be used as an anchor point to trace   the sequence into the past (at higher z) or be extrapolated  into the future (to
lower redshift) relative to the observed z. This is the assumption that we   test in this work, and more specifically we aim to testing whether the chronology of the spatial distribution of ionised gas can be recovered from a 21 cm observation at a single redshift.

%\ddom{It is important to keep in mind that all these images share the same history of reionisation. Thanks to the wealthness such a quantity offers, a single snapshot of the 21 cm signal provides important informations on the state of gas but also on the reionisation process and its whole history.}

To obtain this chronology we can use the  reionisation time field $\TR$ (\Cite{Chardin2019}).
Mapped on 2D images (see Fig. \ref{fig:zreionMAP}), $\TR$ returns the time of reionisation for each pixel of the map 
%\ddom{. This field allows us to get more information,} 
and encodes the complete history of ionisation propagation in a single field. In \Cite{2022A&A...658A.139T}, it was shown how its topology contains a wealth of information on the reionisation process. For example, $\TR$ minima are the seeds of the propagation fronts where presumably the first sources can be found, $\TR$ isocontours track HII bubbles at a given time or its skeleton provides the sites of ionisation front encounters. It also gives information on the influence of radiation sources on each other \Citep{Thelie2022_2}, opening the door to study distant radiative suppression by nearby objects in the environment. More generally, $\TR$ gives information on local reionisations rather than the global reionisation, putting an emphasis on the environmental modulation of the ionisation history. This local modulation of how light is produced and propagates can translate into local variations of star formation suppression (see e.g. \Cite{Ocvirk_2020}) or influence the spatial distribution of low-mass galaxies (see e.g. \Cite{Ocvirk2011}). Galaxies experience a great diversity of reionisation from their point of view (e.g. (\Cite{Aubert_2018}, \Cite{zhu2019}, \Cite{Sorce2022}), and the reionisation time distribution probes this diversity. Other examples of using a similar description include \Cite{Trac2008} on the thermal imprint of local reionisations, \Cite{Trac_2022} for reionisation modelling, or \Cite{Deparis_2019} for ionisation front speed measurments. It should be noted that these specific examples use reionisation redshifts instead of reionisation times; while directly related, we found that times are more easily reconstructed than redshifts for our purposes (see Appendix), and we   focus only on reionisation times in this paper.

\begin{figure*}[h!]
    \centering
    \includegraphics[width=1\textwidth]{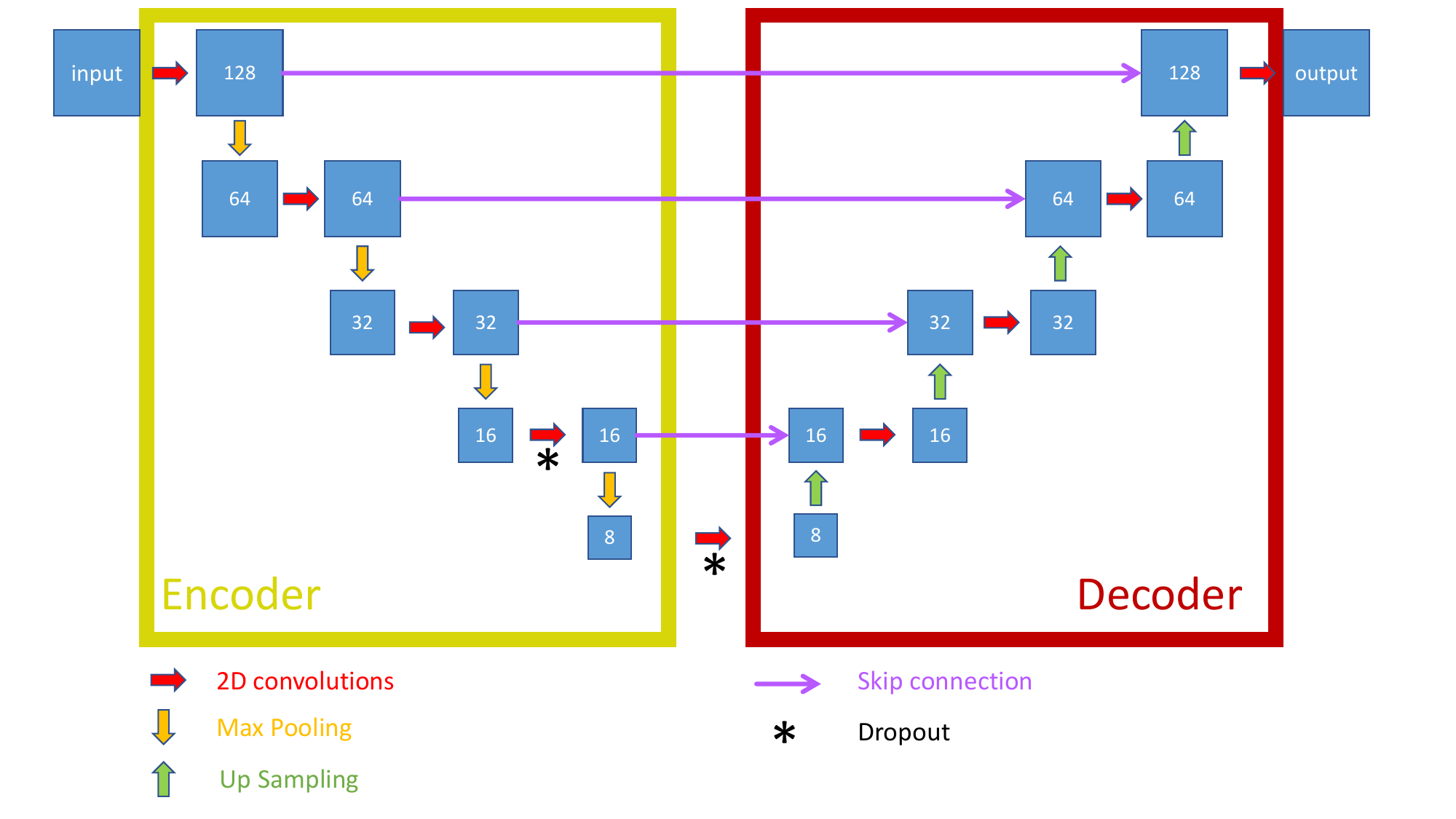}
    \caption{Convolutional Neural Network used in this work. The Encoder (left) refers to the first sequence of 2D convolutions and Max Pooling operations, while the  Decoder (right) refers to the second sequence of 2D convolutions and upsampling operations. Each Max Pooling operation reduces the input map size by a factor of 2, keeping mainly large-scale information. The upsampling operator does the opposite, propagating the information at larger scales and increasing the input map size. Each convolution modifies the number of filters in such a way that we have the maximum number of filters at the deepest point of the network. Dropout and skip connection layers are depicted with black asterisks and purple arrows, respectively. Each square contains the size of the feature maps, which  remain consistent along a given horizontal line (see Appendix \ref{app:CNNalgo} for additional details about the CNN model).}
    \label{fig:unet}
\end{figure*}

As a means to predict $\TR$,  we   use 
%\ddom{machine learning methods.Deep learning (DL) and}
convolutional neural network (CNN) methods, which are capable of detecting and learning complex patterns in images. This tool
%\ddom{, especially in the field of data analysis} 
has been widely used in different problems of astrophysics and cosmology (e.g.  \Cite{Bianco2021}, \Cite{Gillet2019}, \Cite{Chardin2019}, \Cite{Prelogovic2021}, \Cite{Ullmo}). In a recent study \Cite{Korber_2023} successfully retrieve the  growth history of bubbles using mock physical fields. In this study, we extend the CNN applications to $\TR$ field reconstructions from mock observations of the 21 cm signal using a 
%\ddom{deep learning architecture name the} 
U-shaped convolutional neural network (\Cite{Ronneberger2015}), 
%\ddom{In summary, getting $\ZR$ from 21 cm maps allows }
which allowed us to get the whole history of reionisation of a sky patch from a single observation. 

This article is structured as follows. In Sect. \ref{sec:NeuralNetwork} the CNN algorithm and the procedure to deal with the analysis are described. We also present the simulations used to obtain the data. In Sects. \ref{sec:metrics} and \ref{sec:structure} we   present   the metrics used and the results obtained from  monitoring the neural network performance. We  discuss instrumental effects in Sect. \ref{sec:discussion}, and  conclude in Sect. \ref{sec:conclusion}.
%\LEt{***be careful with "eventually": eventually=sure to happen, but at an unspecified later time. It's different from eventuellement (French) or eventualmente (Italian)=can possibly happen under the correct circumstances. Please check throughout.}

 \section{Convolutional neural network and simulation}
\label{sec:NeuralNetwork}

The main purpose of this study is to reconstruct the spatial
distribution of the  reionisation times  from 21 cm images using a convolutional neural network (CNN). CNNs are often used to process pixel data, and became widely used for image recognition (\Cite{LeCun1999}). 
Our neural network is implemented thanks to the  Tensorflow (\Cite{tensorflow2015-whitepaper}) and Keras (\Cite{chollet2015}) Python libraries. It took root in the well-known U-net network first developed by \Cite{Ronneberger2015}. The particularity of this network architecture lies in two distinct parts (Fig. \ref{fig:unet}). The first part  is a contracting path called the encoder, applying series of 2D convolutions and downsamplings to the input image (a 21 cm map here) where its size shrinks as it goes deeper through the neural network. 
Then the second part does the opposite; it  consists of an expansive path (the decoder) applying the same number of convolutions with upsamplings to propagate the information obtained in the encoder. The resulting final output is then another image, $\TR$ in our case. This special case of CNN is called an auto-encoder.

For the learning process, we generated a dataset of histories of reionisation, with their corresponding sequence of 21 cm maps. One CNN predictor was  considered for each $\ZT$ redshift at which we have mock 21 cm observations. 
%\ddom{These redshift is referred to as the observed redshift $\ZT$ and will depict the redshift of 21 cm maps used in the training phase referencing a given model for the network.} 
In practice, we considered 18 predictors for each $\ZT$ shown in Fig. \ref{fig:TB}. Ideally, all CNN predictors   create the same $\TR$ map from mock observations drawn from the same reionisation history. However, depending on the specific properties of a given 21 cm observation (e.g. the non-zero signal fraction) at a given $\ZT$, the predictions will not perform equally well. %It is worth noting that the 18 21 cm maps can be potentially used to provide temporal information of the same region of the sky to our CNN. However, this approach is not feasible in practice since observing the 21 cm sky at a given frequency only provides a region of the sky at a specific depth (redshift), and changing the observed frequency results in another region in the sky at a different time. Therefore, the strategy is to reconstruct the complete history of reionisation based on a single observation at a particular cosmological time.}

The public $\cmfast$ simulation code (\Cite{Mesinger2011}, \Cite{Murray2020}) was chosen to obtain the dataset (i.e. 21 cm signal and $\TR$ fields). Coeval simulations cubes of size 256 cMpch$^{-1}$ with resolution 1cMpch$^{-1}$/pixel were produced using a $\Lambda$CDM cosmology with ($\Omega_m,\Omega_b,\Omega_{\Lambda},h,\sigma_8,n_s$) = (0.31,0.05,0.69,0.68,0.81,0.97) consistent with the results from \Cite{planck2018} and using standard $(T_{vir},\zeta)$ parameters. The parameter $T_{vir}$ sets the minimal virial temperature for halos to enable star formation (see \Cite{theseNico}, \Cite{JulianB2020}, \Cite{BARKANA2001125}) and \Cite{SP.Oh}) and was chosen such that log$_{10}(T_{vir})$=4.69798. The parameter $\zeta$ sets the ionising efficiency of high-z galaxies, and allows us to modify the reionisation timing: the larger this value is, the faster  the reionisation process will be (\Cite{Greig&Mesinger2015}). We   considered two ionising efficiencies $\zeta=30$ and $\zeta=55$ (referred as $\zeta$30 and $\zeta$55), leading to  a total of 36 CNN models to be trained, 18 redshifts per $\zeta$ value.
\begin{figure*}
    \centering
    \includegraphics[width=1\textwidth]{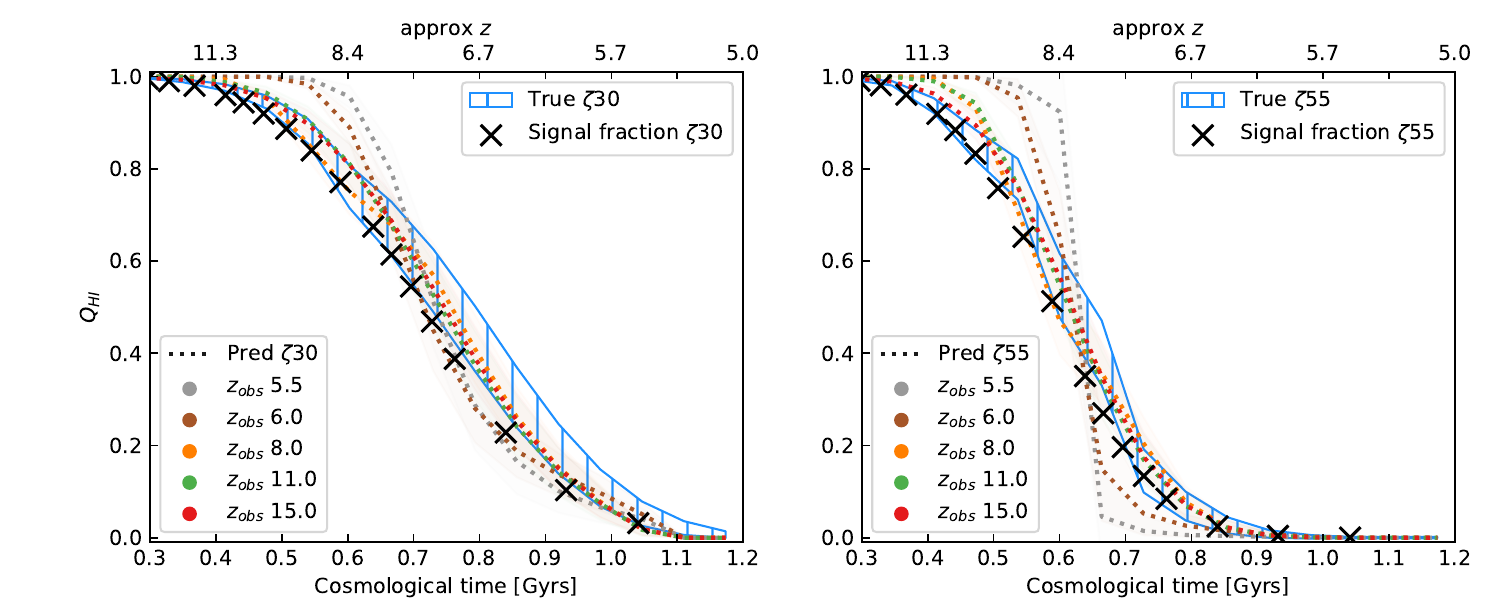}
    \caption{Neutral fraction of hydrogen with respect to time (bottom axis) and redshift (top axis) for $\zeta30$ (left panel) and $\zeta55$ (right panel) . The crosses stand for the average non-zero signal fraction obtained from 21 cm images at different values of  z (as if taken along the line of sight). The dashed areas stand for the true $Q_{HI}$ computed from true maps of $\TR$ (mean and std). The dotted lines stand for the average $Q_{HI}$ obtained from maps of $\TR$ predicted by the CNN.}
    \label{fig:QHII}
\end{figure*}
For each $\zeta$, 50 different realisations with different seeds were run, 
%\ddom{the density of matter, temperature of the gas, and all other cosmological fields.}
giving us access to $\TR$ and 21 cm 3D fields.
As discussed in the introduction, an alternative approach is to consider the reionisation redshift $z_{reion}$(r) instead of $\TR$. However,  we found that the times were better reconstructed, and a brief analysis using $z_{reion}$(r) is presented in Appendix \ref{app:zreion}.
%For the latter, and in order to compare the performance of the algorithm according to $\ZT$, 18 redshifts have been chosen within the range [5.5->15] to produce 21 cm signal cubes. 

To produce 2D images of $\TR$ and the  21 cm signal, we took 64 evenly spaced slices, one out of four (of 1 cMpch$^{-1}$ thickness, corresponding to one cell), in the three directions of each cube.
%\ddom{Furthermore, the \cmfast simulation code use periodic boundary conditions to simulate the behaviour of matter at the edge of the box. This kind of patterns might be learnt by the deep learning algorithm and eventually infer some bias. That's why } 
Each slice was cut into four 128x128 images, finally leading to a total of 768 21 cm images per realisation and per z, giving 38,400 maps per redshifts. We standardised the 21 cm images to ensure that the range of pixel values was consistent across all images in the dataset and to help the model's training process. The mean value was subtracted and the result was divided by the standard deviation (std), both computed over the training set. The mean and standard deviation values are thus the  parameters of our predictors.

As shown in Fig. \ref{fig:QHII}, the neutral volume fraction $Q_{HI}$ 
%\ddom{(being 1-$Q_{HII}$)} 
is shifted (on the time axis) according to the ionising efficiency $\zeta$. Since $\zeta$ controls how many photons escape from galaxies, $\zeta30$ gives a delayed history of reionisation compared to $\zeta55$.   
We discuss below the non-zero signal fraction  (i.e. the fraction of pixels with non-zero 21 cm signal). Its time evolution is plotted as dots and crosses in Fig. \ref{fig:QHII}, and is shown to follow $Q_{HI}$.

%Indeed, taking observation maps at different redshift (line of sight), a value of "signal fraction" can be defined. Doing this at several redshifts, we obtain the non-zero signal fraction on the redshift range, giving dot and cross markers on  

The entire dataset is split into three subsets, from which 35,000 images are used for the learning phase. The first subset, known as the training set, comprises 31,500 images. At each epoch during the learning phase this set is fed to the CNN, which computes the loss function  (i.e. mean square error,  MSE in our case) and modifies the weights to minimise it.  Another separate subset of the entire dataset, called the validation set, consists of 3,500 images;  it is exclusively used to evaluate the CNN's performance during the learning phase after each epoch, without being used in the weight adjustment process.
%zsepasses through the algorithm that will as well calculate the loss function. But it will not update its weights, allowing us to know which value of the loss the neural network will get by predicting images it has never seen. 
The final set is called the test set, and consists of the remaining 3,400 images, which are never processed by the CNN during the learning stages.
%Also, only the 21 cm maps of this set are seen by the CNN algorithm. Hence, both loss function and R$^2$ (see section \ref{sec:internalmetrics}) coefficient will not be calculated on this set. 
All the results shown in this paper (except for the loss function and R$^2$; see Sect.  \ref{sec:internalmetrics})  were obtained via the test set.

\section{Monitoring the algorithm performance}
\label{sec:metrics}

Once all the hyper-parameters were set and predictions made, we     measured the training performance and the prediction accuracy by comparing the predicted $\TR$ maps with the ground truth given by the \cmfast simulation. 

\subsection{Network internal metrics}
%\LEt{***Internal metrics of the network }}
\label{sec:internalmetrics}
First, two internal metrics were used to monitor the training process. Starting with the loss function, the mean square error (MSE) was defined as the average of the squares of the errors. At each epoch the algorithm tries to minimise this loss function (MSE) by comparing the ground truth (given by the simulation) with the prediction (given by the CNN). \\
A second indicator was used, called the determination coefficient,   defined as

%\frac{\Sigma (Pred - \overline{True})^2}{\Sigma (True - \overline{True})^2}
\begin{equation}
   R^2 = 1 - \frac{\Sigma (\mathrm{Pred} - \mathrm{True})^2}{\Sigma (\mathrm{True} - \mathrm{\overline{True}})^2} =  1 - \frac{\Sigma_{n=1}^{N_{pix}} (\mathrm{Pred_n} - \mathrm{True_n})^2}{\Sigma_{n=1}^{N_{pix}} (\mathrm{True_n})^2}
,\end{equation}
where %Where the left expression stands for the general one whereas the right one has been reviewed to fit our case. 
$\mathrm{Pred}$ and $\mathrm{True}$ are the $\TR$ maps of the prediction and ground truth, respectively; $\mathrm{Pred_n}$ and $\mathrm{True_n}$  
%\LEt{***ratio here? or $\mathrm{Pred_n}$ and $\mathrm{True_n}$? (also below) Slashes are used in equations, and to denote ratios, instrument pairings, and wavelength ranges (e.g. optical/UV). All other appearances should be removed and the sentence rephrased. You can substitute "and", "or", "and/or", or a double hyphen (which can be used to indicate a range or dual nature: Hertzprung--Russell diagram). Please check throughout. For more details, see Sect. 2.9 of the language guide  }
correspond to the $n$-{th} pixel of the considered batch of images; and $\mathrm{\overline{True}}$ depicts the average of the true field and is equal to zero after normalisation. In our case the predicted values, $\mathrm{Pred}$, and the ground truth values, $\mathrm{True}$, are both (3500, 128, 128) cubes. The summations are performed over $N_{pix}$ pixels  to measure the network performance on a set it has already or never seen (training or validation set, respectively). Identical values of  $\mathrm{PRED}$ and $\mathrm{TRUE}$ lead to $R^2$=1.

\begin{figure}
    \centering
    \includegraphics[width=0.5\textwidth]{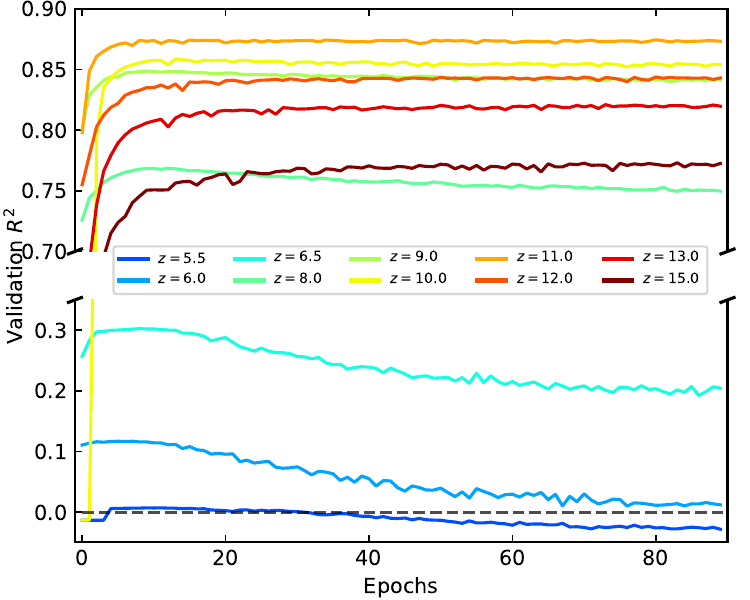}
    \caption{$R^2$ coefficient with respect to epochs for a selection of training redshifts. These curves correspond to the validation phase and are for the $\zeta 55$ model. The curves for $\zeta 30$ are not shown here, but a comparison between the two models is shown in Fig. \ref{fig:R2max}.}
    \label{fig:R2}
\end{figure}

Figure \ref{fig:R2} shows the $R^2$ coefficient during the validation phase for several observation redshifts $\ZT$ and for $\zeta 55$. For the validation set, this coefficient gives a first estimation of the similarities between true fields and predicted fields immediately at the end of each epoch, allowing us to follow the model's accuracy during the learning process. At low redshift the $R^2$ coefficient is small and, even worse, negative for this model and for the lowest z. Nothing can be predicted from low redshift ranges since the non-zero signal fraction at small $\ZT\in[5,7]$ is low or even equal to zero. Then, above a value of  8 we observe that the prediction performance increases for increasing $\ZT$ until 11, where it   eventually degrades again until 15. 

\begin{figure}
    \centering
    \includegraphics[width=0.5\textwidth]{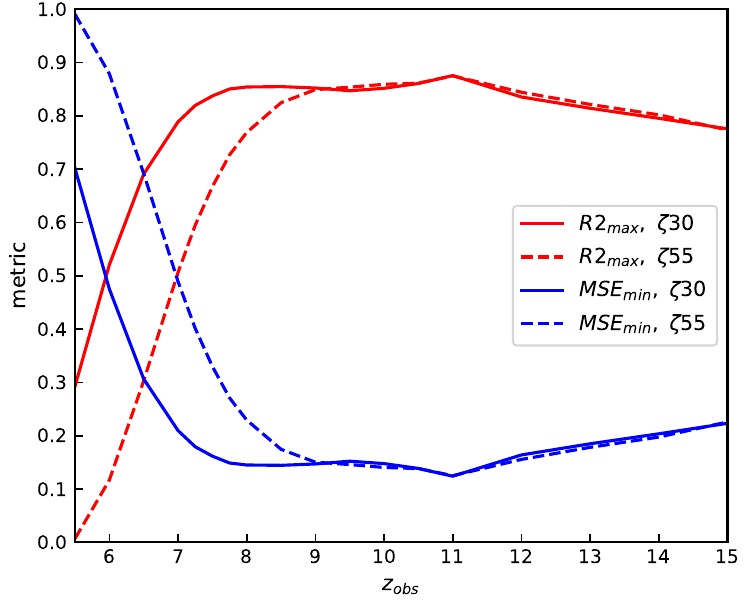}
    \caption{Metric maximum and minimum with respect to the training redshift for the validation set. The red line is for the coefficient R2 and the blue line is for the MSE. The solid lines are for $\zeta55$ and the dashed line for $\zeta 30$ model.}
    \label{fig:R2max}
\end{figure}

\begin{figure*}
    \centering
    \includegraphics[width=1\textwidth]{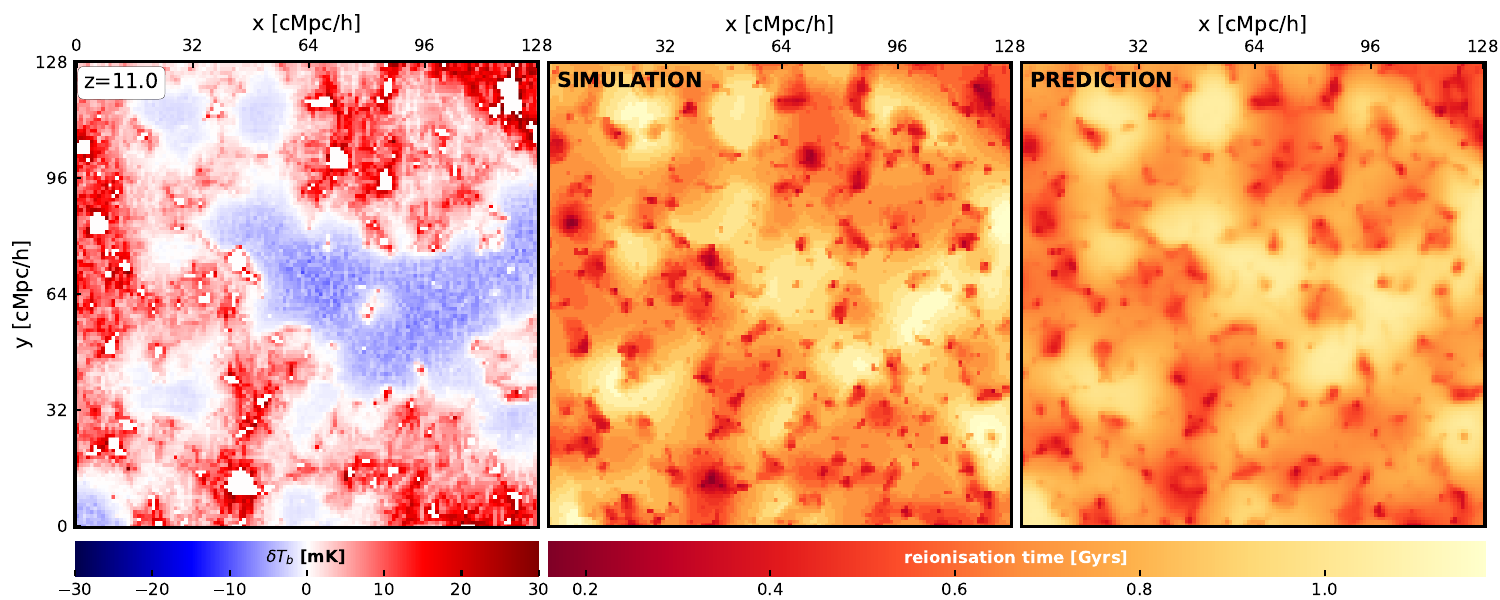}
    \caption{Example of prediction done by the CNN trained from images at $\ZT$=11 and for $\zeta 30$ model. The left panel shows the 21 cm signal $\delta T_b$ at this redshift. The middle panel is the ground truth of the $t_{reion}$ field. The right panel is the CNN prediction for $t_{reion}$ obtained from $\delta T_b$ (left panel).}
    \label{fig:predictions}
\end{figure*}

Figure \ref{fig:R2max} shows the maximum value across epochs for $R^2$ for each $\ZT$ and the two $\zeta$ models, as well as the minimum value reached by the MSE loss. According to these metrics, the best reproduction of $\TR$ is obtained from the CNN predictor using 21 cm maps at $\ZT$=11, corresponding to 95$\%$ of non-zero signal (see Fig. \ref{fig:QHII}). 
Furthermore, the $\zeta 30$ model returns better results at lower $\ZT$: 
at $\ZT$=7, $R^2_{max}$$\approx$0.79 for $\zeta30$, while $R^2_{max}$$\approx$0.51 for $\zeta55$. This can be easily understood from Fig. \ref{fig:QHII}, where the non-zero signal fraction of $\zeta 30$ is considerably larger than for $\zeta 55$ in the $\ZT$  range [5,8] (top axis): at $\ZT$=7, $Q_{HI}$>40$\%$ for $\zeta 30$, while <17$\%$ for $\zeta 55$.
At this range of signal fraction values, the gain in terms of information is such that the network performance increases significantly. From Fig. \ref{fig:QHII} and Fig. \ref{fig:R2max}, we can estimate that between [0.90,0.96] of non-zero signal fraction the neural network achieves better performance.  For a non-zero signal fraction greater than 0.96, the performance decreases again. At these levels of non-zero signal fraction there are only a few HII bubbles to be found, inducing a loss of information on the location of the seeds of most reionisation regions. Without HII bubbles, the sources of reionisation cannot be located and the UV radiation propagation cannot be determined.
%nor how the UV radiation propagates. 
%Nevertheless, there is some rare bubble of HII corresponding to the first source of reionisation positions.
Hence, in order to get the best performance, the CNN algorithm requires a compromise between a minimal set of HII bubbles and a significant non-zero signal fraction. 
{The peak value of $R^2$=0.88} at z=11 also suggests that this redshift of observation is peculiar. The timeline in Fig. \ref{fig:TB} shows that   z=11 seems to be the transition between a global negative temperature brightness and a positive one in our model as the long range influence of X-rays on the gas becomes effective. At this $\ZT$ the 21 cm map contains small HII regions with no signal, easily interpretable for the CNN as the places where the first seeds of reionisation are found. Then there are regions that are hotter than average (shown in red) that will reionise sooner, and blue regions that are  colder than average that will be the last regions to reionise. This $\ZT$ thus contains information of the sequence of radiation propagation that seem to be more easily extracted compared to other observation redshifts.

\begin{figure*}
    \centering
    \includegraphics[width=1\textwidth]{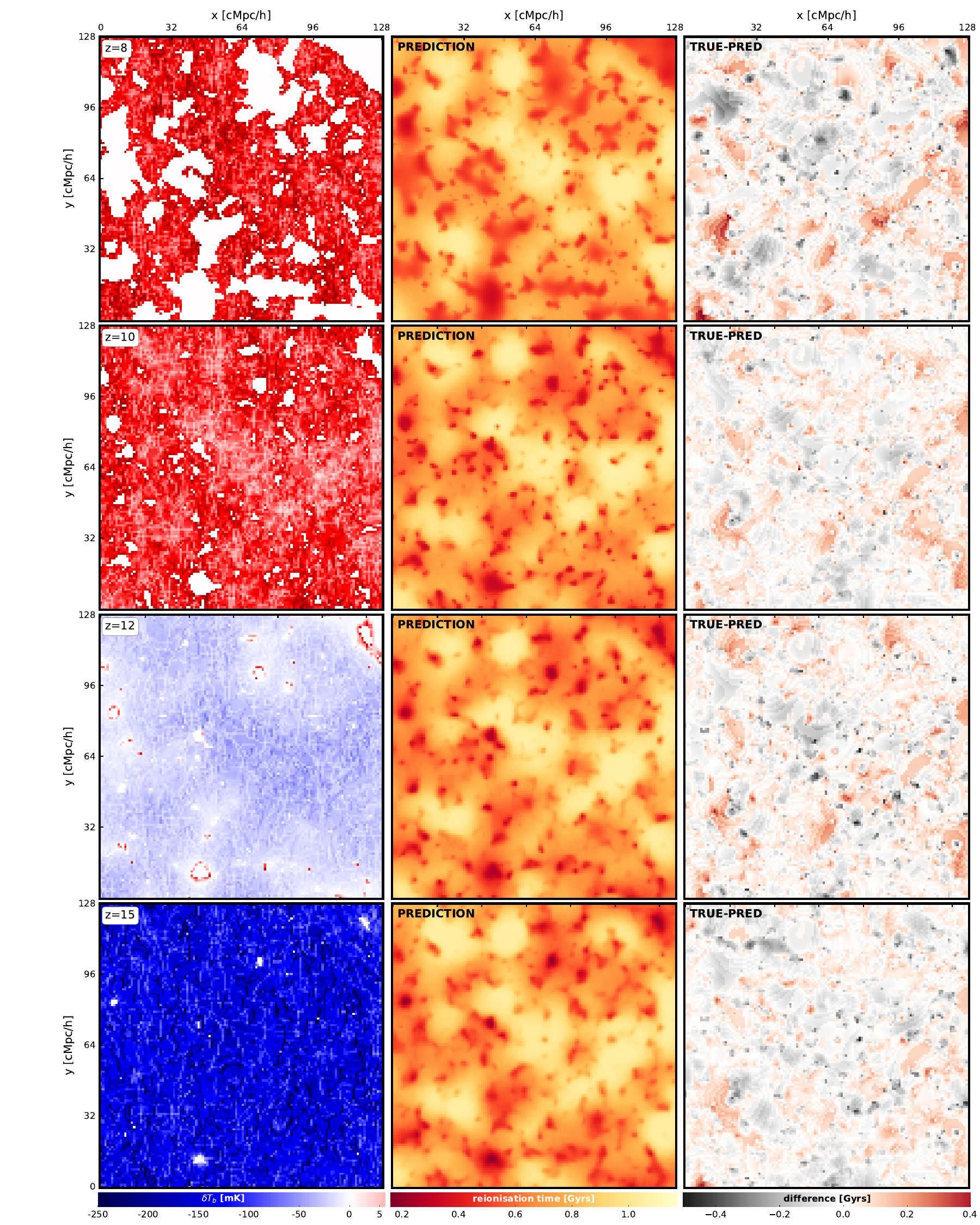}
    \caption{Examples of prediction made by the CNN trained with maps at several redshifts $\ZT$ and $\zeta 30$. The left panels show the  21 cm $\delta T_b$ maps. The middle column shows the predicted $t_{reion}$ fields. The right panels show  the difference TRUE-PRED. Hence, the redder it appears on the map, the more the CNN overestimates the true value at the given pixel; instead,  the darker it is, the more the CNN underestimates the real value. The true $\TR$ is shown in Fig. \ref{fig:predictions}.}
    \label{fig:finalpredictions}
\end{figure*}

\begin{figure}
    \centering
    \includegraphics[width=0.5\textwidth]{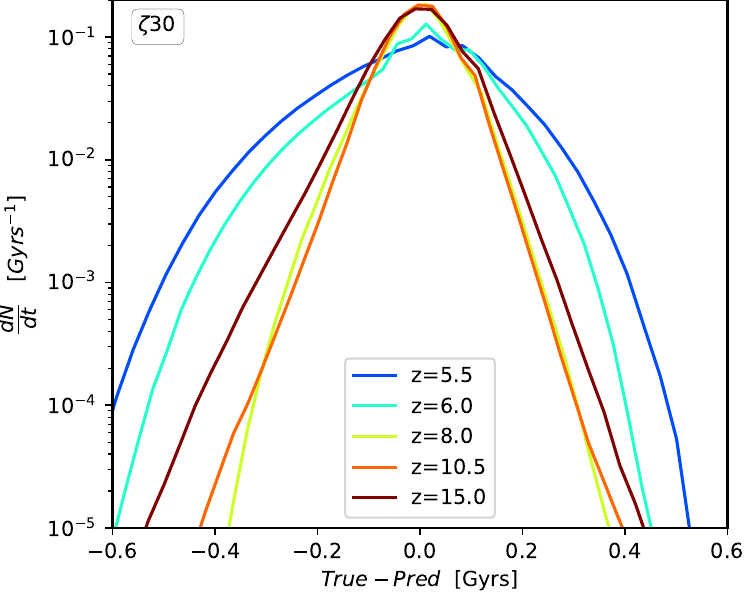}
    \caption{Normalised distribution of TRUE-PRED values (as shown in Fig. \ref{fig:finalpredictions}) for several $\ZT$. These curves were obtained from the whole test set of   $\zeta30$ models.}
    \label{fig:dndt}
\end{figure}

\begin{figure*}
    \centering
    \includegraphics[width=1\textwidth]{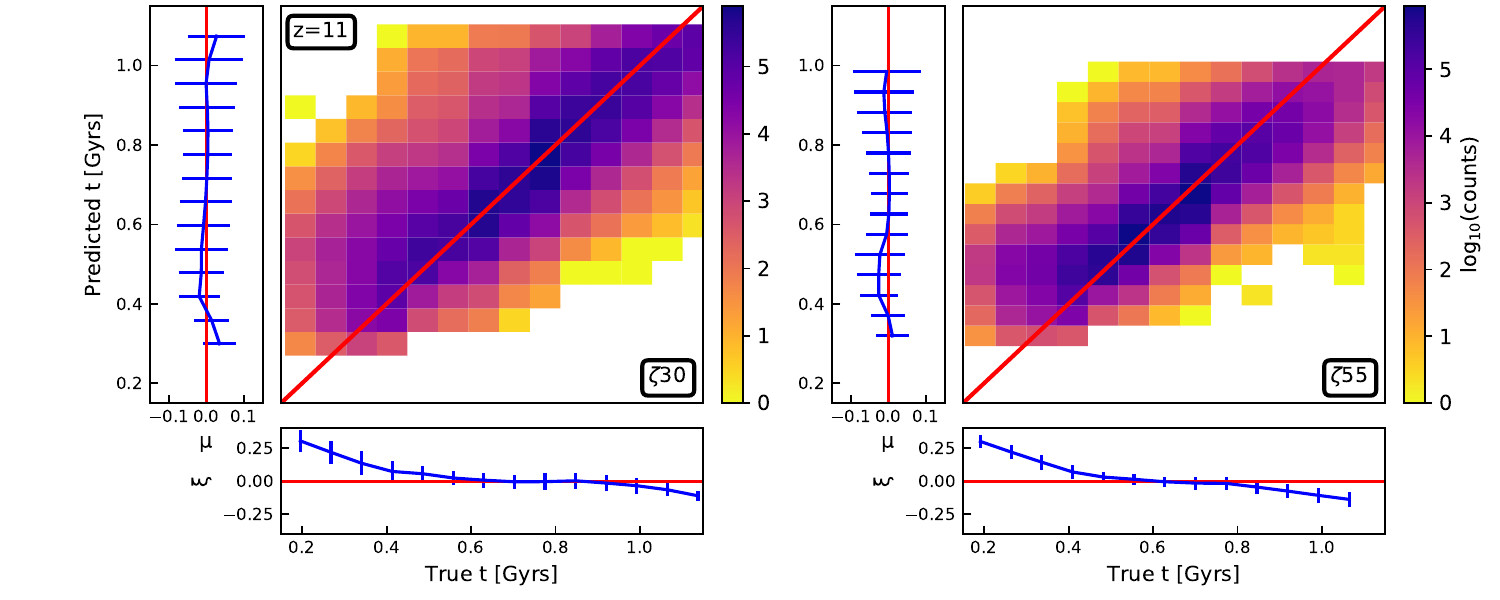}
    \caption{True vs predicted maps obtained from the full dataset in the test set, for $\ZT$ =11 and for both models. The red lines stand for the perfect correlation. The bottom and left histograms are the mean and the standard deviation of the residual r: r = Predicted$-$True, in both vertical and horizontal directions. The bottom histogram stands for the learning error, while the side histogram stands for the recovery uncertainty. In practice, the recovery uncertainty 
    %\LEt{***the side histogram? the recovery uncertainty? }
    is the only accessible estimator since the ground truth will not be accessible.}
    \label{fig:tvp}
\end{figure*}

\begin{figure}
    \centering
    \includegraphics[width=0.5\textwidth]{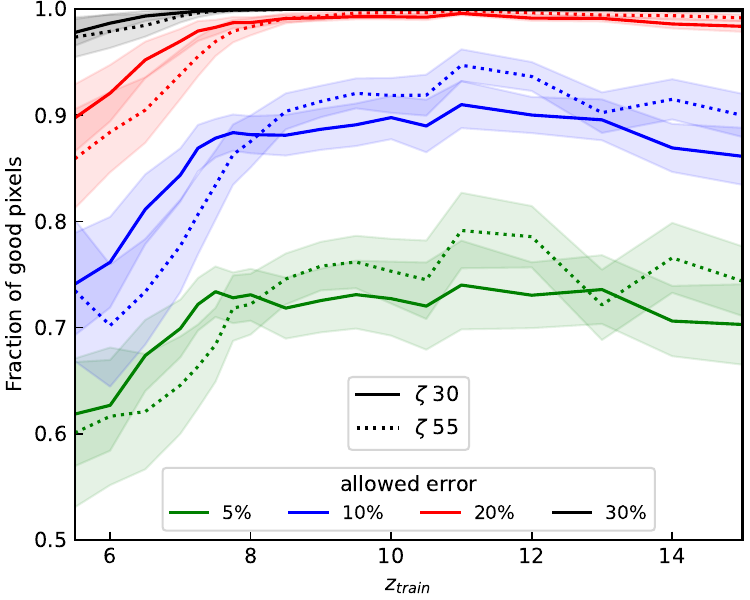}
    \caption{Fitting fraction with respect to redshift. The greater   the allowed error for a pixel (between prediction and ground truth) is, the larger  the number of pixels that have a value close to the true one within this error (i.e. `good pixels'). The solid  lines are for $\zeta 55$ and the dotted lines for $\zeta 30$. The shaded areas depict the standard deviation around the mean for the whole predicted dataset.}
    \label{fig:fitfrac}
\end{figure}

\subsection{Reionisation time prediction}

Beyond the CNN internal metrics, the immediate result is the predicted map itself, as shown in Fig. \ref{fig:predictions}. This $\ZT$=11 map is one of the best reconstructions ($R^2 = 0.84$) we could create for the $\zeta 30$ model. The predicted map on the right seems quite close to the ground truth, but smoother. We note that the best predicted maps for $\zeta 30$ and the best predicted maps for $\zeta 55$ (not shown here) present a similar qualitative behaviour.

With the true map of $\TR$ and its prediction, we can count the number of pixels with values larger than a given reionisation time to obtain Q$_{HI}(t)$ on the sky (see Fig. \ref{fig:QHII}). 
For $\ZT$=11, both TRUE and PRED measurements match within the dispersion of true values and are consistent with the signal fraction evolution computed from the actual evolution of the 21 cm signal with z. This implies that the information obtained across the sky at a single $\ZT$ via the predicted $\TR$ is consistent with (or can be cross-checked against)   the evolution along the line of sight.
%\ddom{The curves in dotted red lines depicts the same quantity but for predicted $Q_{HI}$}. \ddom{For $\ZT$=11, the prediction fits accurately the ground truth. We can then cross check the information obtained in the line of sight and obtained from an observation in the sky.}

Figure \ref{fig:finalpredictions} shows examples of predictions obtained for different models trained at different $\ZT$. The first column shows the mock observations (21 cm maps) at several redshifts, the middle column shows the predictions obtained with the left panel and the right column shows the difference between the ground truth and the predicted $\TR$ field. Looking at the two first columns, at low redshift, the predictions are suboptimal as the inferred field has been completely smoothed. For $\ZT$ $\ge 10$, Our CNN becomes able to capture small-scale features (<10cMpch$^{-1}$),  such as extrema.
%which leads to a more detailed prediction. 
However, in the right column the CNN seems to have more difficulties  in predicting the local extrema of reionisation times, even though their locations are well predicted, especially at low redshift (e.g. for $\ZT$=8: (x,y)$\approx$(120,125)cMpc/h). These points correspond to the seeds of the propagation of fronts, presumably linked to the first sources of radiation, and seem to be subject to 
%\LEt{***to undergo? "suffer" suggests sentience }
a smoothing intrinsic to our CNN implementation. Compared to $\ZT$=10 or 11, the $\ZT$ =15 prediction appears to be slighty smoother, although the earliest reionisation times seem to be well reproduced.
%It is easily understood since 21 cm maps outside $\ZT$<5.5 contain the very first bubbles of ionised hydrogen.
Finally, Fig. \ref{fig:dndt} depicts the normalised histogram of TRUE-PRED maps.
Distributions are centred on zero, with an assymetry  towards negative values: our CNN predictions returns greater reionisation times than the ground truth (i.e. a delayed reionisation history). For example, taking $\ZT$=15 there are more pixels at True-Pred=-0.4 Gyr (dN/dt>2e-4) than for True-Pred=0.4 Gyr (dN/dt<3e-5). This systematic effect is less severe for the best CNN predictors trained to process $\ZT$=8 or 10.5 observations in this figure.
%The majority of value turns around a difference of 0 Gyr, yet slitghly asymmetrical, the distribution being larger on the left. Meaning that for the full range of $\ZT$, the CNN tends to give a delayed reionisation history compared with the ground truth. 

%In summary, depending on $\ZT$, predictions can be more or less accurate. In the best case, it's only smoothing further the map, keeping large scale structures and losing small scale information. In the worst case, the predictions are totally wrong and especially at low redshift (between z=5 and z=7). While the redshift range of [8,12] seems to be the best compromise to get the best results, higher $\ZT$ such as 15 give a better representation of the first sources of reionisation.

\subsection{True versus Predicted histograms  and fitting fraction}

One of the most standard tests is the   true versus predicted (TvP), where all the predicted pixels are compared one by one to their true value given by the simulation.
%\LEt{***One-sentence paragraphs should not be used. Please check throughout, and include single sentences in the previous or following paragraph, as appropriate  }\\
Figure \ref{fig:tvp} shows the TvP corresponding to all the maps in the test set using the $\ZT$=11 CNN. Most values follow the perfect correlation for typical $\TR$ values (0.4-1 Gyr), while extreme values (<0.4 Gyr and >1 Gyr) are not as well recovered by the CNN (mean value up to +0.25 and down to -0.15 respectively). This is not surprising, given the predicted maps on Fig. \ref{fig:predictions}.
The extreme values of $\TR$ %are precisely where the algorithm fails at having a perfect prediction, they 
coincide with small-scale features that are smoothed out, where the first sources with lowest $t_{reion}$ are found. These values are also rare (4.2$\%$ of the total number of pixels), explaining why the algorithm fails at learning how to recover them. 
%It will tend to fill at a lower/higher value that is more predominant explaining why extreme values are overestimated/underestimated. 
%It results on a smoother map that keep the mean reionisation time at a relevant value close to the mean value of the ground truth.

Figure \ref{fig:fitfrac} 
%\ddom{tries to give additional information in addition to}
presents a synthesis of the true versus  predicted maps of all the predictors at different $\ZT$. The fitting fraction is a value between 0 and 1, which  corresponds to the predicted pixel's fraction whose value fits within an arbitrary error calculated as $\epsilon\%$ of the true pixel's value. 
%\ddom{Each pixel is compared (pred/true) and is counted 1 if the predicted pixel does not exceed the allowed error.} 
The larger the allowed error is, the more `good' pixels will be found. It is then clear how low $\ZT$ values (<8) gives less accurate results, especially when allowing a small error (<10$\%$): the fitting fraction value is more than 10$\%$ lower than for $\ZT$>8. It can be understood from the maps of temperature brightness (or from Fig. \ref{fig:QHII}) where the maps contain less   signal with decreasing redshifts from z=8: less than $50\%$ of the map contains observable neutral hydrogen. At the extreme redshift z=5.5 there is no signal left because all HI is  reionised and no prediction is possible. A value of  %However the CNN tries to predict the $\TR$ field anyway, putting a mean value close to the ground truth, explaining why the accuracy at the lowest $\ZT$ is not zero.
$\ZT$>8 seems to give the best results, and all predictions made from $\ZT$ above 9 seem to have similar performance: between 70 and 75$\%$ of matching pixels for 5$\%$ error, and a slight decrease with growing $\ZT$. Overall, we recover the two trends identified previously: low $\ZT$  simply lacks the signal for a good reconstruction, while high $\ZT$ lacks the direct imprint of sources that appear later. The best compromise is found for $\ZT$ values between 8 and 12, corresponding to signal fraction $\approx$ 0.8, with an optimal value of  0.95. 
%To sum up this first step study briefly, low $\ZT$ is nonviable to predict $\TR$ fields due to a lack of information. Meanwhile high redshift maps also deteriorate the prediction, but more fairly for the same reason as stated previously in sec \ref{sec:internalmetrics}, the CNN algorithm needs the first HII bubbles to efficiently locate the first sources of reionisation. Then, maps containing around 80$\%$ ($\ZT$=10, see Fig.\ref{fig:QHII}) non-zero signals seem to be rather good to predict $\ZR$ field while around 95$\%$ ($\ZT$=12) of signal fraction seems optimal.

\subsection{$\zeta30$ and $\zeta55$}
The first results comparing the  $\zeta30$ and $\zeta55$ scenarios are very close. The prediction accuracy is similar except for lower $\ZT$ (<8) where the signal fraction tends to be quite different between both situations. In the most extreme case, at $\ZT$=5.5, the CNN trained with the  $\zeta$30 maps is quite limited, but still returns a prediction, whereas the CNN trained with $\zeta$55 maps cannot predict anything, which  was expected since there is no HI left at late times in the $\zeta55$ scenario. In the following sections, we   only consider the $\zeta30$  scenario to discuss results in the whole range of $\ZT$ used.
\begin{figure}
    \centering
    \includegraphics[width=0.5\textwidth]{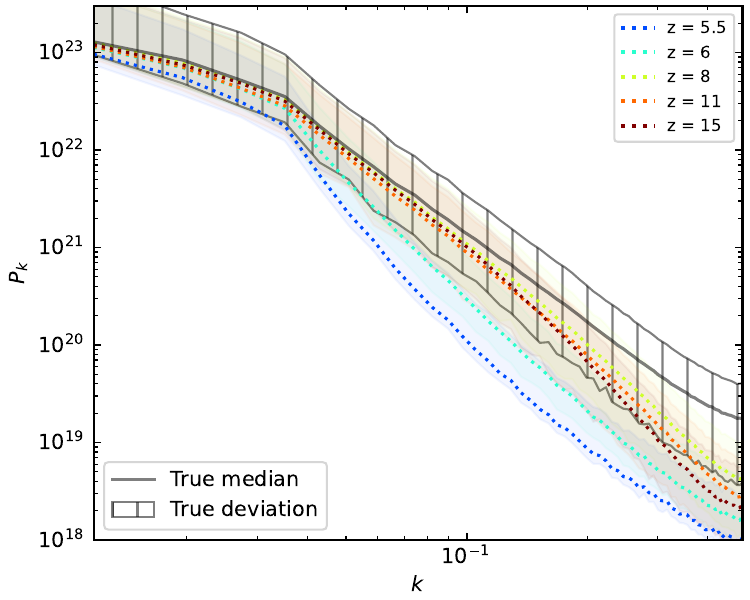}
    \caption{Power spectrum of the $\TR$ field obtained for $\zeta 30$. k is the spatial frequency with dimension [cMpc$^{-1}$h]. The black line stands for the average power spectrum of the ground truth and the dotted coloured lines stand for $P(k)$ predictions with different $\ZT$. The shaded areas depict the dispersion around the median (1${st}$ and 99${th}$ percentiles) for both predictions and ground truth.}
    \label{fig:Pk30}
\end{figure}

\section{Structure of predicted maps}
\label{sec:structure}

We now investigate the spatial structure of the reconstructions using three metrics: the power spectrum, the Dice coefficient, and the minima statistics.

\subsection{Power spectrum $P_k$}

%$R^2$/MSE metrics only give a global vision of the algorithm performance. TvP directly compare pixels together. It will not tell us how well the neural network is doing at a specific scale.
We now compare the power spectrum $P_k$ of the reionisation time field with that predicted by the neural network in order to have a statistical point of view on how well the network reconstructs the different scales present on the map.
%\LEt{***paragraph }
Figure \ref{fig:Pk30} depicts the $\TR$ power spectrum of model $\zeta$30. A first look shows that the lack of 21 cm signal drastically erases the possibility to predict anything. Predictions for z=5.5 and z=6 are incompatible with the real $P_k$ at mid-scales k $\approx$ 7e-2 cMpc$^{-1}$h; less than 30$\%$ of the power remains for z=6, against up to 85$\%$ for z=8. At small scales (k > 2e-1 cMpc$^{-1}$h) less than 12$\%$ of the power remains for z =6 against up to 57$\%$ for z=8. Now looking at large scales (low spatial frequencies such as k < 3e-2 cMpc$^{-1}$h), our model reproduces the power at more than 95$\%$ for $\ZT>8$. However, at k=0.2 cMpc$^{-1}$h and beyond, 
%, corresponding to a resolution of 5 cMpc/h, i.e. 5 pixels, 
the prediction cannot produce enough power, meaning that the smaller scales are difficult to predict at all $\ZT$. Again, the predictor smoothes the $\TR$ field, predicting a map generally blurrier than the ground truth. 
%\ddom{For large scales studies, this information about the property of the $\ZR$ power spectrum can be enough to statistically describe the history of reionisation.}
To improve results at the smallest scales, generative adversarial networks (GANs) 
%\ddom{, a special architecture of Deep learning networks,} 
could be a solution (see \Cite{Ullmo}). 

\subsection{Dice coefficient}
\begin{figure*}
    \centering
    \includegraphics[width=0.8\textwidth]{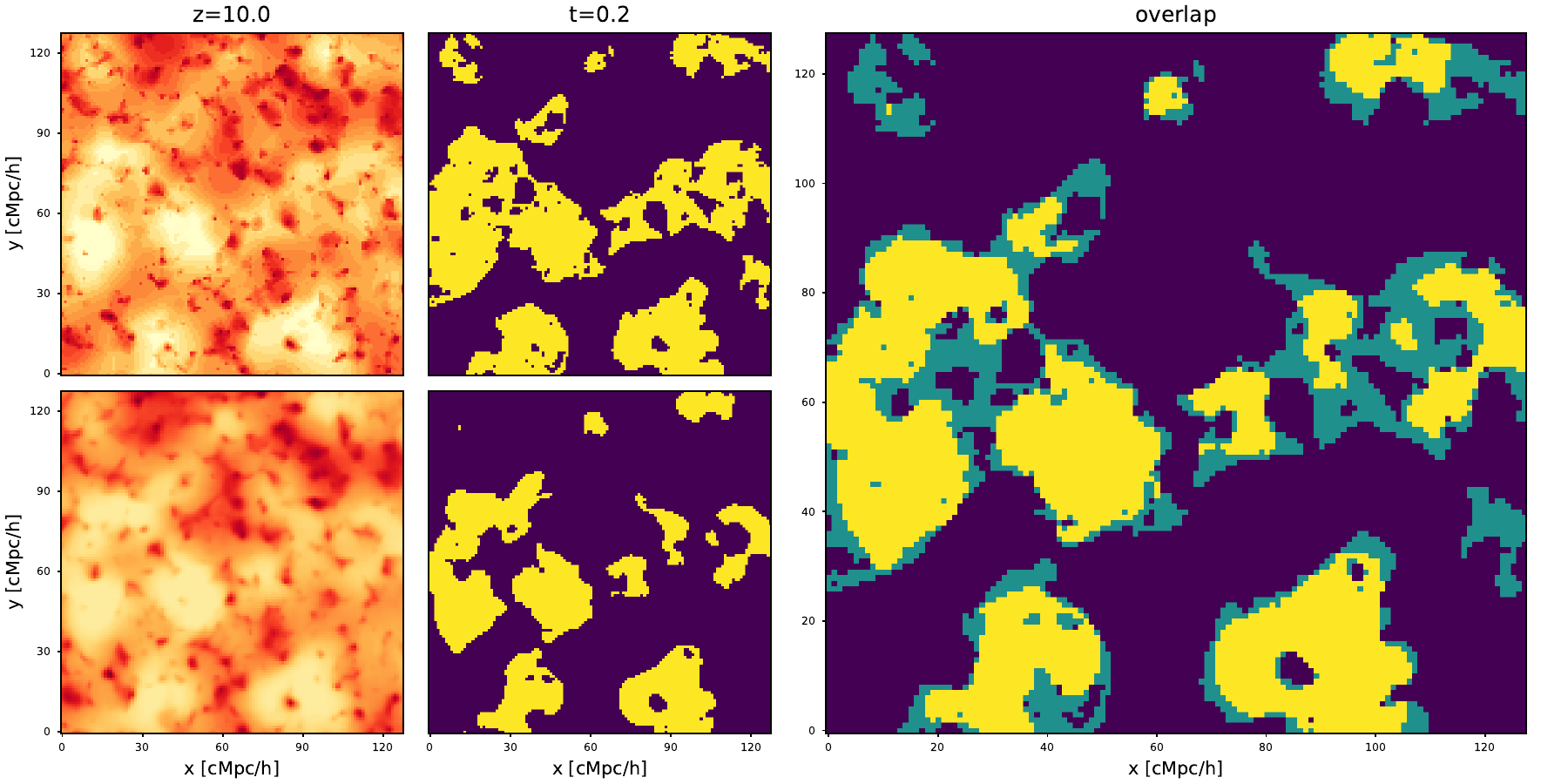}
    \caption{Example of overlap map for $\zeta 30$ and a threshold value $t=0.2$. The left panel is the true $t_{reion}$ field (top) and the predicted $\TR$ for a $\ZT$ of 10 (bottom). The middle panel depicts a 1 or 0 map where 1 (yellow) stands for the pixels fitting 20 \%\ of the highest values and 0 (purple) for pixels below the threshold for the  true field (top) and the predicted field (bottom). The right picture is the overlap of the middle panels, where yellow stands for coincidence, green for incorrect pixels in the prediction, and purple where the values are below the threshold in both fields.}
    \label{fig:overlap}
\end{figure*}
Another way to look at the predictor performance is the Dice coefficient (see \Cite{Ullmo}). This method is useful to see what kind of regions the algorithm reconstructs   best; for example, whether the first regions that become reionised are clearly predicted or if, conversely, late regions are reconstructed in a better way. This coefficient focuses on the map structure by looking at regions with given values. It   determines 
%\LEt{***yes? "tell" suggests ability to speak) }
how the CNN recovers structures instead of giving an accuracy according to the value of pixels or the considered scale.

The Dice coefficient is used by taking a threshold t (0 to 100) and by considering only the   pixels with the largest $t_{reion}$ values in the true and predicted maps. We can estimate the  regions of the map where the prediction overlaps with the ground truth, using a newly formed map with pixels in only three possible states:
\begin{itemize}[label=\textbullet, font=\small \color{black}]
    \item Predicted and true pixels  both have a value above  the threshold (both fit
the criterion), denoted  the yellow state.
    \item Both have a value below  the threshold (both do not fit the criterion), denoted the blue state.
    \item There is a mismatch between the prediction and simulation, denoted the green state.
\end{itemize}

\begin{figure}
    \centering
    \includegraphics[width=0.5\textwidth]{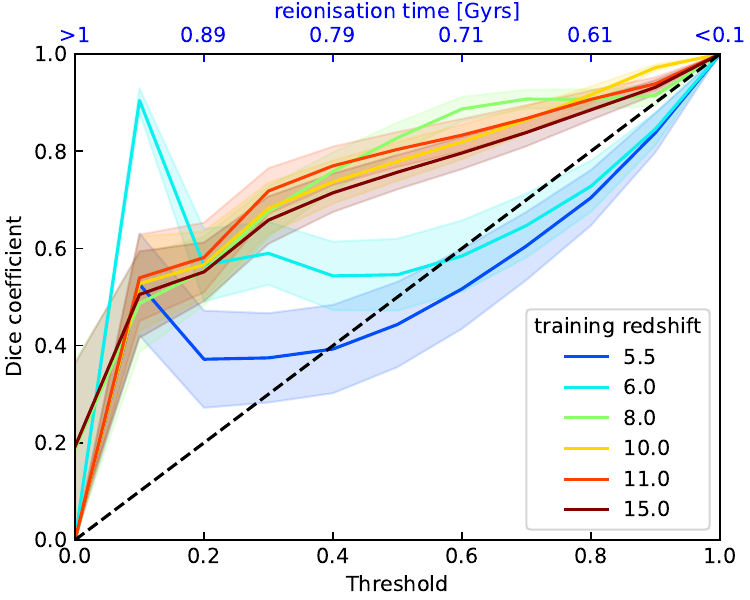}
    \caption{Dice coefficient for several $\ZT$ and for $\zeta30$.  The dashed black line stands for the Dice coefficient when matching the true map with a random mask. Pixels are considered starting with the largest pixel value (last pixel to reionise) down to a threshold. A 0.2 threshold means that we consider the last 20\% to be reionised. 
    %The threshold depicts how many pixels we count, starting with the highest pixel values (last pixels to reionise). The last pixel value taken gives then the time of reionisation at the given threshold. 
    The shaded areas depict the standard deviation around the mean DICE coefficient.
    }
    \label{fig:Dice30}
\end{figure}

It is  important   to note   that the value of the threshold corresponds to a given cosmological time (or redshift). Using 10$\%$ for the threshold, the constructed map will only contain the information for large values of cosmological time (low redshifts), typically the last regions to reionise. On the other hand, taking 100$\%$ as threshold, the whole map will be considered.
%since we ask for reionised regions belonging to the full range of cosmological time. 
%Coming back to Fig. \ref{fig:overlap}, the network trained with images of $\delta T_b$ at $\ZT$=10 gives good results. 
An example of an overlap map is depicted in Fig. \ref{fig:overlap} using $\ZT$=10. The threshold example on the figure is 0.4, corresponding to 40$\%$ of the largest values. Only a few green regions (corresponding to 15$\%$ of the pixels) are present, and more than 85$\%$ of the pixels are in agreement between the prediction and the ground truth. This range of values is actually well reconstructed, and the remaining differences are located at the edge of these regions (in the green areas, e.g. (x,y)$\approx$(70,10)cMpc/h).

The Dice coefficient, or association index, is calculated at a given threshold as \citep{DICE}
\begin{equation}
  Dice = \frac{n_{yellow}}{n_{yellow}+n_{green}},
\end{equation}
with $n_i$ the number of pixels with colour i. The Dice coefficient can only take values between 0 and 1 (0 for no correspondence between prediction and ground truth and 1 for a perfect reconstruction).

Figure \ref{fig:Dice30} shows the Dice coefficient for the $\zeta30$ model.
%according to threshold values and for a set of $\ZT$. 
%\ddom{Taking the left bottom corner of each panels, threshold t=0$\%$, meaning we only take top 0$\%$ pixels. Obviously  no pixels fit within the condition. At the opposite, in the top right corner, all the pixels fit in the condition and are taken.} 
The dotted black line
%\ddom{line gives us a reference. It} 
stands for the Dice coefficient obtained if we compare the true thresholded map with a map randomly filled with zeros and ones. Globally, we recover that $\ZT$>8 provides better accuracy compared to the random situation (for threshold=0.4: Dice>0.7 against 0.4, respectively), with similar performance  at all $\ZT$. Furthermore, $\ZT$=11 seems to   dominate until median values for the threshold. Afterwards, $\ZT$=8 coefficents catch up, followed by $\ZT$=10, meaning that these observation redshifts recover efficiently the first structures of reionisation.

Looking at low threshold values, the Dice coefficient provides additional insights into the performance at low $\ZT$, such as $\ZT$=5.5 or 6. At these redshifts, predictions are slightly better for low threshold values meaning that at $z_{obs}$=6, the neural network predicts in a very efficient way the last regions reionised  (for threshold=0.1, Dice>0.9): since they are the only regions where a non-zero signal can be found, the predictor can locate them accurately. %It is not that surprising since observations at z=6 and lower redshift only contain a low amount of information (low signal fraction, see Fig. \ref{fig:QHII}) and then the global efficiency remains low. Indeed, the neutral bubbles of hydrogen found in these observations are precisely regions of the universe not reionised yet, it is then easier for the predictor to locate them accurately. As the threshold increases, the efficiency of these low $\ZT$ decreases drastically.

%Eventually for large value of threshold (firsts sources of reionisation) and for $\ZT$<6, the CNN is less effective than for a random distribution. Indeed, as already stated, maps at $\ZT$<6 are almost fully empty giving no solution for the predictor. 
%\ddom{However, going to the right in the x-axis, the DICE coefficient shut down quickly and drastically since these redshifts of observation do not contain enough information to efficiently predict the whole field. 
%Eventually, if one wants to focus on the last regions to reionise, it could be enough to look at low redshift observations.}

\subsection{Minima statistics}

\begin{figure}
    \centering
    \includegraphics[width=0.5\textwidth]{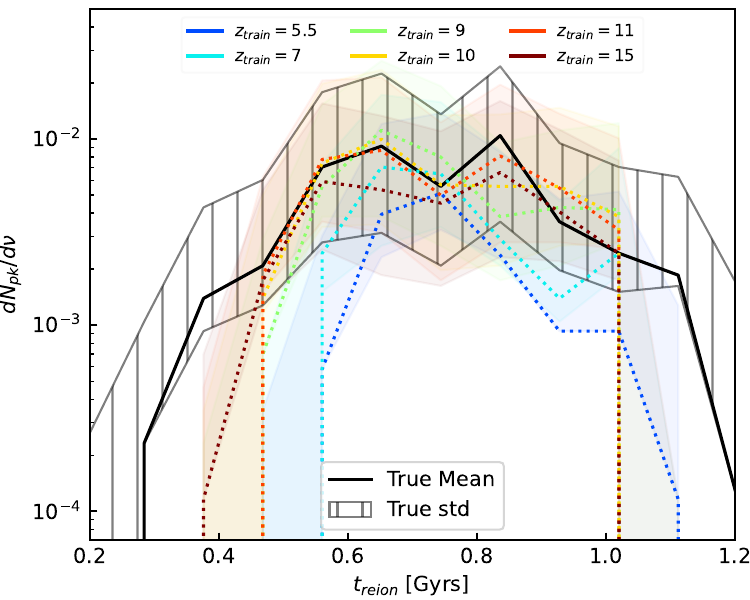}
    \caption{Distribution of reionisation times for $\TR$ minima. The solid line is obtained from true $\TR,$ while the dotted lines are obtained from predictions at several $\ZT$. The shaded areas show the dispersion around the median (mean and std) for both predictions and ground truth.}
    \label{fig:Peaks}
\end{figure}

\begin{figure*}
    \centering
    \includegraphics[width=1\textwidth]{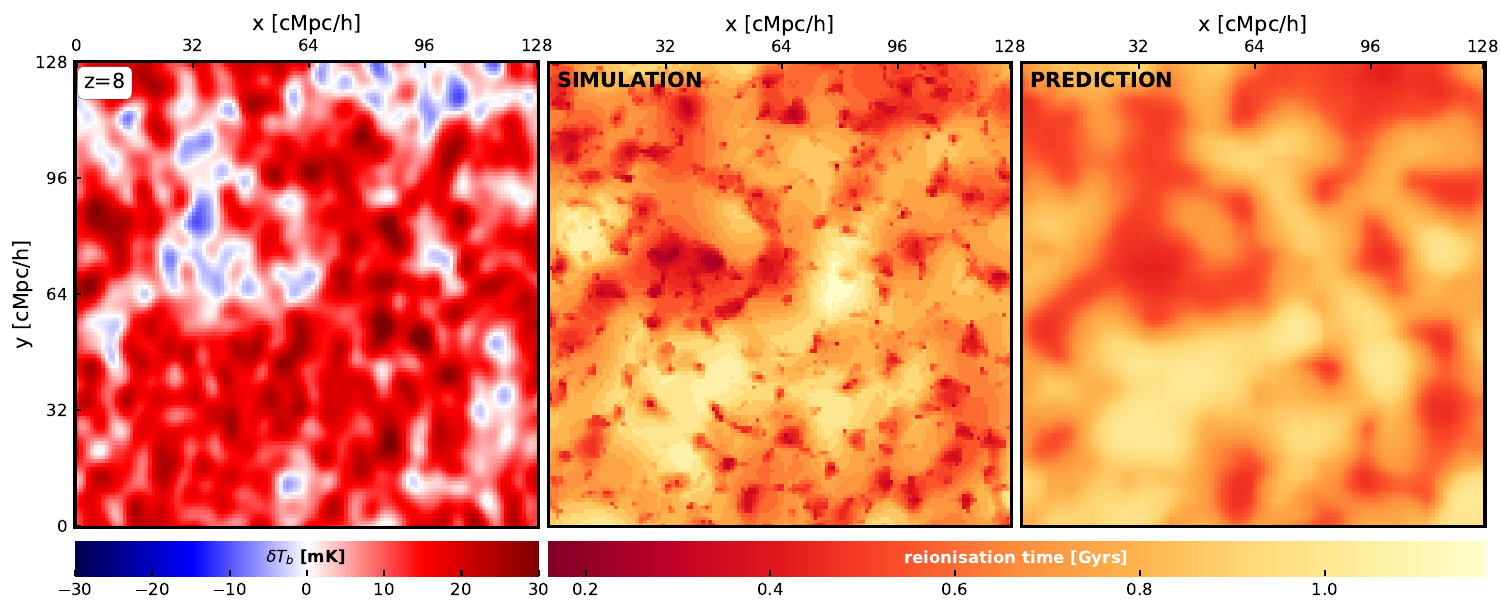}
    \caption{Predictions for the model $\zeta$30 at $\ZT$=8, including instrumental effects for a typical SKA observation (left panel, see details in main text). The middle panel is the ground truth for $\TR$ and the right panel is the CNN prediction.}
    \label{fig:noisypred}
\end{figure*}

\begin{figure}
    \centering   
    \includegraphics[width=0.5\textwidth]{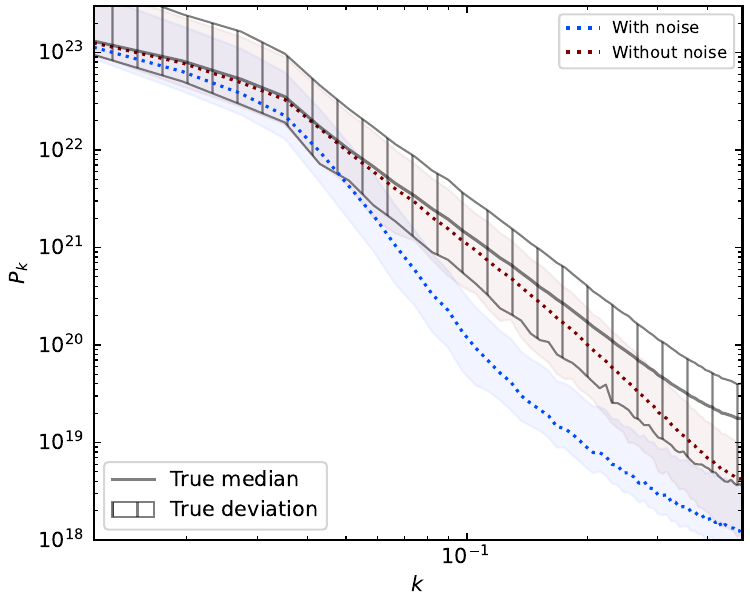}
    \caption{Power spectrum of the $\TR$ field obtained with the model $\zeta30$ at $\ZT$=8, including instrumental effects for a typical SKA observation. The black line and hatched area stand for the mean and standard deviation of the true field $P(k)$. The dotted lines and shaded area depict the median and dispersion ($1{st}$ and $99{th}$ percentiles) of CNN predictions $P(k)$ with (blue) and without (red) instrumental effects.}
    \label{fig:pknoise}
\end{figure}

\begin{figure}
    \centering   
    \includegraphics[width=0.5\textwidth]{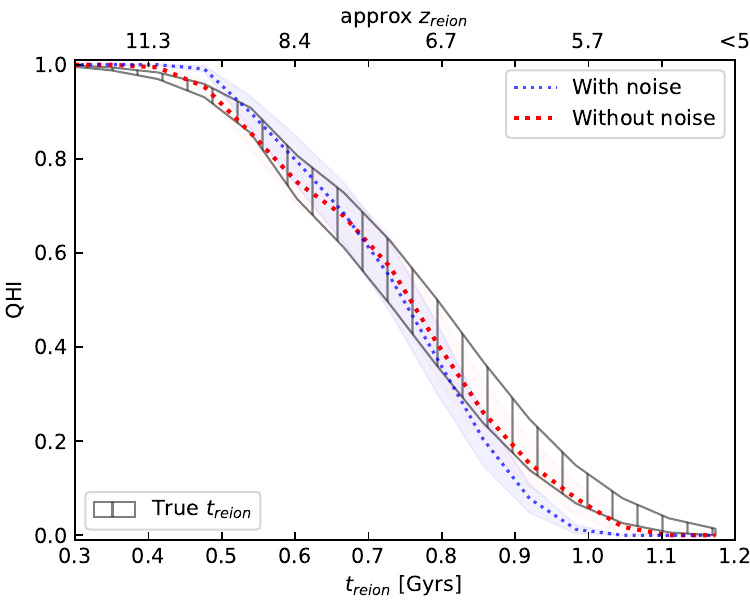}
    \caption{QHI obtained from predictions at $\ZT$=8 with the model $\zeta30$, including instrumental effects for a typical SKA observation. Hatched area stands for the true field QHI. Dotted lines and shaded area depict the mean and standard deviation of CNN predictions QHI with (blue) and without (red) instrumental effects.}
    \label{fig:QHInoise}
\end{figure}

We now investigate at the minima of $\TR$ (i.e. the regions that reionised at the earliest times) to probe how our CNN detects sources of reionisation.
%We use DisPerSE (Discrete Persistent Structures Extractor) (\Cite{Sousbie2011}, \Cite{Sousbie20112}) to identify persistent topological features within 2D maps and to return the distribution of the peak of reionisation times. 
We used DisPerSE (Discrete Persistent Structures Extractor) 
%\LEt{***abbreviation to introduce?}
(\Cite{Sousbie2011}, \Cite{Sousbie20112}) to identify the distribution of reionisation times minima;  it uses discrete Morse theory to identify persistent topological features in two-dimensional maps, such as voids, walls, filaments, and clusters. While we focus here only on minima, valuable insights into  the underlying topology of $\TR$ can also be obtained from the persistent structures detected by DisPerSE, to see how they relate to the physical processes that shape the distribution of reionisation (see also \Cite{Thelie2022_2} a, b).

The results of this analysis are illustrated in Fig. \ref{fig:Peaks}.
The black line stands for this analysis performed on the true $\TR$ field: at low $t_{reion}$ (high z) sources are rare, and their number is highest at t=0.8 Gyr (approx z=7). Their number then drops at higher $\TR$ values, not because sources becomes rarer but because they appear in already reionised regions and cannot be traced by peaks in $\TR$ maps.
%\ddom{there are only a few amounts of sources that started the reionisation. In the case $\zeta 30$, the first sources appear approximately at z>12. As time passes, more sources emerge until a peak at z=7, telling us that the most number of reionisation sources appear at such an epoch.}
%\ddom{At lower redshift, the number of sources that appear seems to decrease. 
%However, new halos and galaxies helping in the reionisation process continue to emerge, but they are not detected as maxima in $\ZR$ field. It is mainly because these sources light up where the IGM is already fully reionised. That's why on the left of the red line, it drops drastically, no more seed of reionisation front appears from this time even though new sources still emerge.}
The statistics of our CNN predictions, depicted with dotted lines,   clearly show that the CNN has some difficulties to detect the first sources of ionising photons ($\TR$<0.4), resulting from the smoothing of $\TR$ maps. However,  the maximum of the distribution is well matched at least for $\ZT$>8, whereas low $\ZT$=5.5-7 predictors fail unsurprisingly to recover the seeds of ionisation fronts from 21 cm maps with very low non-zero signal fractions. Interestingly, an observation made for example at $\ZT$=10 manages to predict in a satisfying manner, by being consistent with dispersion of time distribution, the population of peaks (and thus seeds and/or sources of reionisation) at later times ($t_{reion}$=1 Gyr  ), emphasising the ability of our CNN to extrapolate the future of a given observation. When compared with the previous power spectrum analysis, these results emphasise that the loss of accuracy on small scales  mostly have  an impact on high-z (low $t_{reion}$) peaks, whereas the seeds of ionisation fronts at lower redshifts are much better predicted with the lowest $\ZT$. 
%\ddom{It results from CNN erasing high redshift values and smoothing the global $\ZR$ maps. It also overestimates the main peak's redshift and eventually gives accurate results for the last sources that appear. The same conclusion as previously can be done concerning the best redshift to observe, under z=7, predictions are not accurate at all. They only find few peaks in a limited range of reionisation redshift. Between z=10 and 15, predictions are nevertheless consistent with the ground truth. For $\ZR$<10 the peak statistic is even well represented and especially for the model $\zeta$30. A delayed history of reionisation is then more favorable for predictions.}

%--------------------------------------------------------------------
%--------------------------------------------------------------------
%--------------------------------------------------------------------

\section{Instrumental effects and prediction}
\label{sec:discussion}

%Adding noise and instrumental effect on the input images grants the advantage to have mock images way closer to what the real SKA observations will be. 
The work discussed previously only takes into account a `perfect' 21 cm  signal, without any noise or instrumental effect. Studying the effect of foreground contamination and how to manage it is of primordial importance to being able to properly analyse the future observation. While this paper does not   delve into these effects, this section aims to shed some light on them. Nonetheless, there are other studies in the field of deep learning that specifically address the impact of foreground contamination, such as the work by \Cite{bianco2023deep}. These effects are expected to degrade the predictor's capability to infer the $\TR$ field. 
%However, a lot of studies have been done concerning noise and the behaviour of artificial intelligence on such images (\Cite{TIANchunwei}, \Cite{inbook}). The results are actually accurate since adding noise in images degrades fairly the neural network prediction. 
As a means   to study the potential impact of these effects in our predictions, we created a new dataset of 21 cm maps with instrumental and noise characteristics corresponding to SKA. The UV 
%\LEt{***UV }
coverage and instrumental effects are calculated using the  tools21cm\footnote{https://github.com/sambit-giri/tools21 cm} library (\Cite{Giri2020}), assuming a daily scan of 6 hours, 10s integration time, for a total observation of 1000 hours  
%\LEt{***perhaps: a daily scan of 6 hours, with 10s integration time, for a total observation of 1000 hours }
(\Cite{2022MNRAS.509.3852P}, \Cite{Ghara_2016}, \Cite{Giri_2018}). Our investigation is limited to $\ZT$=8, corresponding to the lowest redshift where the predictor accuracy in terms of the  $R^2$ coefficient remains satisfying ($R^2$=0.86), while deeper observations are found to be significantly more degraded by noise. A maximum baseline of 2 km is assumed and the angular resolution is $\Delta \theta \sim 2.8 $ arcmin, corresponding to 7.35 cMpc at this redshift. The tools21cm library also convolves the coeval 21 cm cube in the frequency direction with a matching resolution $\Delta \nu \sim 0.43$ MHz.
%. Also as the observation goes deeper (at larger redsfhit), the noise level increase significantly, degrading drastically the prediction capability. Hence the optimal redshift of observation according to our first investigations is 8. 
Figure \ref{fig:noisypred} shows the prediction from noisy observations with the input 21 cm observations shown in the left panel. %It immediately demonstrates the differences between the perfect 21 cm field and the SKA observations. T
The predicted  $\TR$ map is much blurrier  at first sight; adding instrumental effects on input observations smoothes  the prediction even more. 

Figures \ref{fig:pknoise} and  \ref{fig:QHInoise} depict  the power spectrum of the $\TR$ field and $Q_{HI}$ for $\zeta30$, for both ground truth and prediction. The two predicted curves (dotted lines) have been implemented with observation at z=8, one with instrumental noise  (in blue) and the other using a perfect 21 cm map (in red). In both predictions the power spectrum is successfully recovered at large scales (k < 3e-2 cMpc$^{-1}h$), with approximately 70$\%$ and 95$\%$ of the remaining power for the noisy case and perfect case, respectively. At smaller scales (k > 2e-1cMpc$^{-1}h$) the power spectrum recovered from noisy maps has a sharper turn-off (17 $\%$ of power remaining against 86$\%$ for the perfect case) and comes out of the error bars (hatched and shaded areas). It results in missing the small-scale structures, making it difficult and even impossible, to detect the first sources of reionisation accurately. On the other hand, $Q_{HI}$ gives a fair history of reionisation, although more sudden from SKA maps than the ground truth and the perfect signal scenario: noisy reionisation ends around $t_{reion}$=1 Gyr and the perfect scenario ends around 1.1 Gyr.

This outcome raises several questions. The first  is linked to the architecture of the CNN, and whether  it is  possible to improve the CNN such that the accuracy of the prediction improves significantly, especially for noisy observations. 
%\LEt{***direct questions should be avoided in formal papers. I have changed it to an indirect question.  }
For this purpose a solution could be to tune the hyper-parameters in order to find the best combination to recover $\TR$. In addition, keeping the U-shaped CNN  and modifying the number of hidden layers or filters, or the deepness of the algorithm could change the prediction in the right direction. Another possibility could be to add an attention block to help our CNN to focus on small-scale features (\Cite{oktay2018attention}).
%However, CNNs are highly sensitive to change and it could also deteriorate the prediction. 
Using GANs (\Cite{Ullmo}) could also improve the output field in order to recover small scales (k > 2e-1cMpc$^{-1}h$).
Another solution would be to preprocess the noisy 21 cm observation in order to remove or reduce noise and instrumental effects. In such a situation we would hopefully recover a perfect observation scenario, drastically boosting the prediction accuracy.

%Improving small scales predictions is pretty interesting since it would allow us to study parameters that affect mainly small-scale structures. One of them is the Dark Matter (DM) model. Indeed, Working with warm or hot DM (WDM, HDM) affects the properties of the 21 cm map and then to each field associated with it, such as $\TR$. For example, the mass of the DM particle used in the simulation directly affects the density power spectrum at the smallest scales. 
%All quantities (T21 field, $\TR$, etc...) of power spectra are then also affected by such a parameter. This is mainly why small scales are really relevant to study. 
%DM model also affects $Q_{HI}$ in a similar way that $\zeta$ does (delaying or hasting the reionisation history) showing that the reionisation history is highly degenerated. It is then difficult to know the exact model to use while analysing future observations.

%There are also a few  more verification we can do with the predicted maps such as verifying if we could for example, recover topological properties as studied in the paper of \Cite{2022A&A...658A.139T} and see how well the network predicts such aspects. Indeed, topology properties are robust to the resolution, we can then detect them even if a map is smoother than reality.

\section{Discussion and perspectives}

The work presented in this paper has involved numerous decision-making processes that may have been influenced by factors such as default settings, initial ideas, and implementation challenges. 
%\LEt{***paragraph }
%To provide a deeper understanding of the rationale behind these decisions, we aim to address some frequently asked questions and provide the lecturer with informative insights.
When it comes to the architecture and hyper-parameters of the CNN algorithm, the choices made are typically based on the fact that they yield improved performance (in terms of $R^2$). However, some choices, such as the number of filters, the inclusion or location of dropout layers, or the choice of loss function for weight adjustment, can potentially be modified; it is conceivable that untested  combinations of hyper-parameters might yield better results. Ongoing investigations are being conducted to further explore this matter.

Another decision was to use images of the 21 cm signal at a single redshift or frequency, leading to one CNN per redshift (and per model). A possibility would have been to train the predictors with multiple redshifts channels or even light cones. This could possibly help the predictors to infer maps even in the regime of low non-zero signal fraction at low $\ZT$ (<8);  the inclusion of  higher $\ZT$ information in the prediction process can provide additional constraints (e.g. on the global reionisation history) that cannot be inferred from single low $\ZT$ 21cm maps alone. Our choice is largely the result of the history of this work, where it was not obvious at first that any prediction would have been possible, even in the case of a perfect 21cm signal. Investigations are currently ongoing to see what can be gained from a multiple channel prediction.

However, we also believe that having multiple CNNs has some merit regarding the adequacy of the parameters (cosmological, astrophysical) of a predictor to the parameters that drive a given 21cm observation. In a real case scenario, the `real' parameters of an observation are unknown, and we therefore face a situation where it is unclear which CNN should be used to reconstruct $\TR$. One possibility is to assume that the model parameters will be obtained from another analysis (e.g. using  the 21cm power spectrum) and the role of a CNN predictor is therefore limited to reconstructing the spatial distribution for the reionisation times in a specific observation. However,  preliminary investigations also show that when a set of 21cm maps at different $\ZT$ are processed by the multiple predictors of a `wrong' model (for example $\zeta55$ maps in $\zeta30$ predictors), they lead to a set of $\TR$ maps that are inconsistent, for example with regard to their average reionisation history. Meanwhile, a CNN that can reconstruct multiple $\TR$ at once from multiple 21m maps would always, by construction, ensure some consistency between its predictions, even for a wrong model. 
%It implies that a set of CNN at different $\ZT$ provides a mean to quantify autonomously the adequacy of its model to the data. 
The optimal situation is likely to be an intermediate situation with CNNs dedicated to reconstruct a given $\ZT$, but that use 
%\LEt{***yes? (ingest=consume, eat) }
multiple 21cm maps at different redshifts.  

\section{Conclusions}
\label{sec:conclusion}

In this study, we   implemented and tested a U-net architecture to infer the $\TR$  field from 21 cm maps simulated by the \cmfast simulation code. These predictions are especially effective to recover the large-scale features ( > 10 cMpch$^{-1}$)  of reionisation times and can to some extent recover the past and extrapolate the future evolution of an observation made at a given $\ZT$.  %With such information on real observation, we could focus on the sky's region where the very first galaxies and stars have been formed and then try to get their properties.  
%Such work has been possible thanks to a wide mock observation dataset of the well-known 21 cm signal. Common scenarios of the reionisation have been taken in here, however, such an algorithm can be applied to other scenarios changing as many parameters as wanted. It is then robust and will always give coherent results if well-trained. 
%The redshift of the 21 cm observation drastically influences the quality of the reionisation times reconstruction. It seems to have a link between the signal fraction on the map and the accuracy of predictions. The signal fraction is directly linked to the observation's redshift. Hence, in our case, 

For our models, $z_{obs}$ values between 8 and 12 seem to provide the best results according to several metrics (e.g.  $R^2$, the Dice coefficient, power spectrum, true vs predicted), corresponding to a signal fraction of  65$\%$ up to 96$\%$   for the $\zeta$30 model. For $\ZT$<8, even though the last regions to be reionised can be reconstructed, the lack of 21 cm signal in general significantly degrades the network's capability to predict $\TR$.  
%still provides a good reconstruction of the last regions to be reionised.  %until totally destroying possibilities to infer anything. 
%Going too far in the past  seems to degrade as well predictions by smoothing further $\TR$. 
For deep observations ($z_{obs}$>12), the CNN still manages to reproduce quite well the very first sources of reionisation due to the rare and narrow HII bubbles imprinted in the 21 cm signal, but has more difficulties to predict the location of sources that appear later, which leads to smoother maps. 
%It results in a lack of information outside these bubbles explaining a smoother map. 
It also might be interesting to keep the information with low signal fractions ($z_{obs}$<8) since it reconstructs quite well the last regions to become reionised. 

In addition, our CNN model works well at recovering the largest scales, as seen for example in the power spectrum analysis. % It also finds the first regions and the last regions that ionised quite precisely. 
Nevertheless, there are still some limits to what our network can do; for example, it has  difficulties to recover the smallest scales ( < 10 cMpch$^{-1}$). That could indeed be a problem to constrain physics related to small-scale structures (such as the physics of low-mass objects or that related to the nature of dark matter). It might be still possible to improve results at small scales with the use of GANs to generate a more detailed $\TR$ field. In addition, implementing an attention block to insert it in our CNN could help predictors to focus on small-scale features. 

Two scenarios have been used with different histories of reionisation. No significant difference in the training phase or in the prediction phase has been detected. The main difference comes at the lowest redshifts: the $\zeta$55 scenario reionises sooner, lacks signal more rapidly, and is more difficult to predict for low $\ZT$ (<8).

%We want to highly insist on studies already made on noise \Cite{Bianco2021}, \Cite{Prelogovic2021} that are primordial for our science. Thanks to prior studies, we know that neural networks are robust to noises even if it slightly degrades (by smoothing) predictions. At shown in section \ref{sec:discussion} it will deteriorate the accuracy for the smallest scales. At the end of the day, with real observations, we expect to obtain a smoothed/blurrier version of the predictions.

We believe that the method presented here can prove to be useful for the future interpretation of 21 cm data. First, it demonstrates that the information of reionisation times is 
%somehow\LEt{***=in an unspecified way } 
encoded in the 21 cm signal. 
%and can, in principle, be derived from observations even though more investigations are needed for our CNN to be able to cope with instrumental effects. Training a predictor using multi-channels (at several $\ZT$) could also help predictors to improve its performance at lower $\ZT$ (<8).} 
The field $\TR$ gives access to chronology of light propagation in the transverse plane of the sky, that could be for example cross-checked with other estimates of the reionisation evolution obtained along the line of sight (21 cm light cones or Ly$\alpha$/21 cm forests for example), and presumably it can be related to the global history of structure build-up and star formation. Another application would be the cross-correlation of reionisation time maps with galaxy distributions or intensity maps other than 21 cm; having access to the propagation history around objects observed through other channels could provide insights into their own local history of the production of  light  (and therefore of stars and sources)  (see also e.g. \Cite{Aubert_2018},  \Cite{Sorce2022}). There should also be an environmental modulation of star formation suppression by reionisation (e.g. \Cite{Ocvirk_2020},\Cite{Ocvirk2011}) and a map of reionisation times  could provide a way to test this by providing an insight into how local reionisation proceeded. As illustrated in Sect. \ref{sec:discussion}, the reconstruction of reionisation times from actual 21 cm data will be certainly be challenging, but surely holds some potential that we have not fully investigated yet.

%Notably thinking about the Hydrogen Epoch of reionisation array (HERA) or SKA radio telescope that is still in construction and will provide observations in the late 2020s. Developing and improving such tools may turn out to be particularly powerful to get ready to analyze the large number of data we will get. 

%Future works will allow us to push even more the U-net architecture at its limit with noisy observation and extra reionisation parameters in order to help us to constrain fundamental keys of the EoR.

\section*{Acknowledgement}
The authors would like to acknowledge the High Performance Computing Centre of the University of Strasbourg for supporting this work by providing scientific support and access to computing resources. Part of the computing resources were funded by the Equipex Equip@Meso project (Programme Investissements d’Avenir) and the CPER Alsacalcul/Big Data. This work was granted access to the HPC resources of TGCC under the allocations 2023-A0130411049 “Simulation des signaux et processus de l’aube cosmique et Réionisation de l’Univers” made by GENCI.
%This research made use of tools21cm, a community-developed package for the analysis of the 21 cm signals from the EoR and Cosmic Dawn (Giri et al. 2020), astropy, a community-developed core Python package for astronomy (Astropy Collaboration 2018); matplotlib, a Python library for publication quality graphics (Hunter 2007); scipy, a Pythonbased ecosystem of open-source software for mathematics, science, and engineering (Virtanen et al. 2020); numpy (Harris et al. 2020) and Ipython (Perez & Granger 2007)}
We also thank J. Freundlich for his help and advice. The authors acknowledge funding from the European Research Council (ERC) under the European Unions Horizon 2020 research and innovation programme (grant agreement No. 834148).

\bibliographystyle{aa} % style aa.bst
\bibliography{biblio.bib} % your references Yourfile.bib

%--------------------------------------------------------------------
%--------------------------------------------------------------------
%--------------------------------------------------------------------

\newpage
\begin{appendix}

\section{CNN architecture details and hyper-parameters}
\label{app:CNNalgo}

\begin{table*}
%\centering
\caption{Details of the architecture the CNN used to predict maps of $\TR$.}
 \begin{tabular}{||c|c c c c c||} 
 \hline
 Network branch & Layer number & Layer type & Nbr of filters/data & Size of filter/data & Activation function\\ [0.5ex] 
 \hline\hline
 \multirow{5}{4em}{Encoder}
 & 1 & Input & 31 500 & 128x128 & . \\ 
 & 2 & Conv2D+Max Pooling & 32 & 3x3 & Relu \\
 & 3 & Conv2D+Max Pooling & 64 & 3x3 & Relu \\
 & 4 & Conv2D+Max Pooling & 128 & 3x3 & Relu \\
 & 5 & Conv2D+Max Pooling+Dropout& 256 & 3x3 & Relu \\
 & 6 & Conv2D+Dropout& 512 & 3x3 & Relu \\ %[0.5ex]
 \hline\hline
 \multirow{5}{4em}{Decoder}
 & 7 & UpSampling+Merge+Conv2D & 256 & 3x3 & Relu \\
 & 8 & UpSampling+Merge+Conv2D & 128 & 3x3 & Relu \\ 
 & 9 & UpSampling+Merge+Conv2D & 64 & 3x3 & Relu \\ 
 & 10 & UpSampling+Merge+Conv2D & 32 & 3x3 & Linear \\ 
 & 11 & Output & 31 500 & 128x128 & . \\ 
 \hline
 \end{tabular}
Each convolution layer within the encoder part is followed by a Max Pooling layer, except the sixth. Instead, each convolution layer within the decoder part is followed by an up-sampling layer plus a Merge layer that concatenates layers of same dimension of the encoder part with the corresponding layer of the decoder.
 \label{Unet_detail}
\end{table*}

In this section we  describe the  details of the CNN algorithm used in this study. We used the Python libraries Tensorflow (\Cite{tensorflow2015-whitepaper}) and Keras (\Cite{chollet2015}) to implement our CNN. Table \ref{Unet_detail} shows properties of the hidden layers. 
A convolution layer consists in applying a filter of a given size (3x3 in this study) to the whole input resulting in a featured map. Each convolution is performed with same padding, meaning that each convolution conserves the size of the input. 
For the encoder part, the   first four convolutions are followed by a Max Pooling 
%\LEt{***sometimes you use MaxPooling and sometimes max-pooling (also Max Pooling within the figures). Can these be  consistent?  }
operation (of size 2x2) that is shrinking the size of the input by a factor of 2. When the 2x2 matrix passes through the input, it only conserves the pixel with the highest value. 
Dropout (Drop) layers are also used in order to help the CNN to prevent overfitting (\Cite{Labach}). A dropout layer acts by shutting down a given fraction (0.5 for us) of neurons/filters in the corresponding hidden layer where it is applied.
For the decoder part, each convolution is followed by an UpSampling layer that doubles the size of the input. 
In addition, a concatenated layer (Merge or skip connection) is applied to fuse features of a given hidden layer within the encoder with features of the same size in the decoder. In practice, skip 
%\LEt{***and make sure all your terminology is consistent as well: e.g. Skip, skip; Dropout, Drop, drop}
connections tend to improve the accuracy of CNN and to make it converge faster (\Cite{XJM}). 
Eventually, the activation function for each layer is Relu, except for the last one  that is a linear activation function since we want to predict an output field with continuous values.
During the learning phase of a CNN algorithm, the weights of each convolution filter are updated each time a batch of data is passed through. In our implementation, we   set the hyper-parameter batch size to 16. This means that our model   updates its weights after processing each batch of 16 images, and that one epoch is completed after N batches have been processed, where N is the number of batches needed to cover the entire dataset.
In addition, to optimise the performance of the  CNN algorithm, we needed to carefully choose the hyper-parameters. Some of them have already been discussed, such as batch size, dropout, and loss function. The `optimizer' hyper-parameter was set to Adam. Another important factor is the initial weight of the model. The `kernel$\_$initializer' hyper-parameter controls this and was set to `He Normal'. However, because the weights are randomly initialised, there is a possibility that the learning process may get stuck in a local minimum without learning any more. To prevent this, we added a feature to the code that restarts the weight initialisation if such a situation is detected.

\section{Using $\ZR$ instead of $\TR$ }
\label{app:zreion}

The redshift of reionisation $\ZR$ and $\TR$ are two fields depicting the same quantity: the time of reionisation of regions in the sky, with $\ZR \sim \TR^{-2/3}$ during the reionisation epoch. %Furthermore, the time $t$ can be expressed as a function of the redshift z such that $t(z)\propto z^{-2/3}$. Such a transformation results on $\TR$ being more gaussian than $\ZR$. 
We can investigate how choosing redshift instead of time affects the capability of the CNN to predict $\ZR$ or $\TR$ and which field gives the best results. 
%In this section, we bring elements of answers that made us turn to $\TR$.

\begin{figure}[hbt!]
    \centering
    \includegraphics[width=0.5\textwidth]{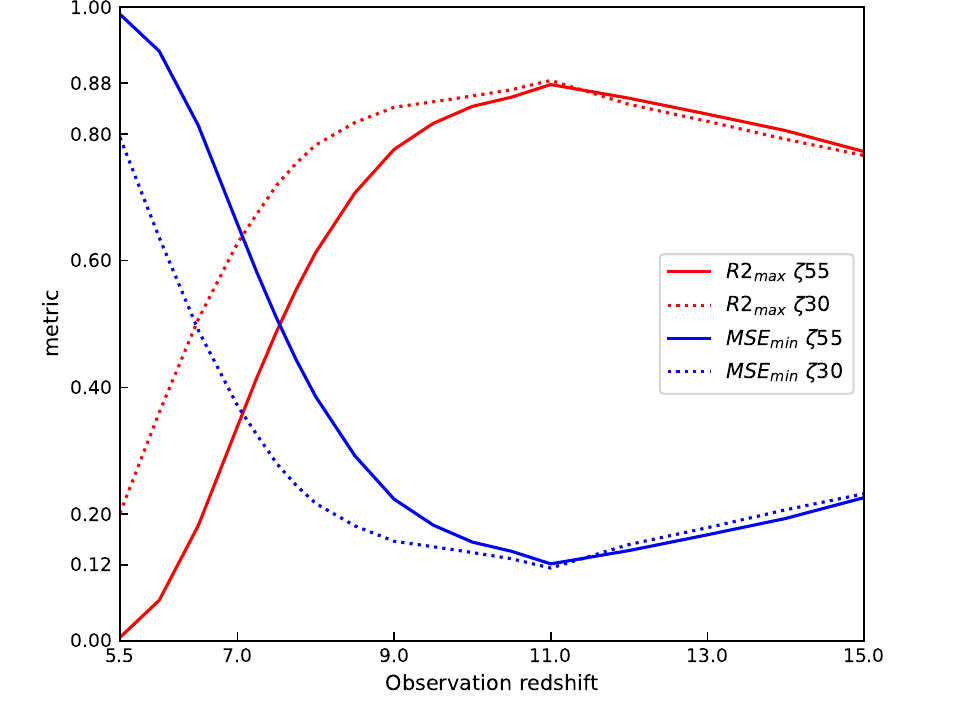}
    \caption{Metric maximum and minimum with respect to the training redshift for the validation set. The red line is for the coefficient R2 and the blue line is for the MSE. The solid  lines are for $\zeta55$ and the dashed line for $\zeta 30$ model.}
    \label{fig:ZR2max}
\end{figure}

First, Fig. \ref{fig:ZR2max} shows the same plot as Fig. \ref{fig:R2max}, but for $\ZR$. For z=11 the values of the R2 and the MSE are quite similar. Nevertheless, going to lower values of z results in both metrics getting worse much faster than for $\TR$. Working with time and not redshift provides a greater range of redshifts for which the results are relevant. 

\begin{figure*}
    \centering
    \includegraphics[width=1\textwidth]{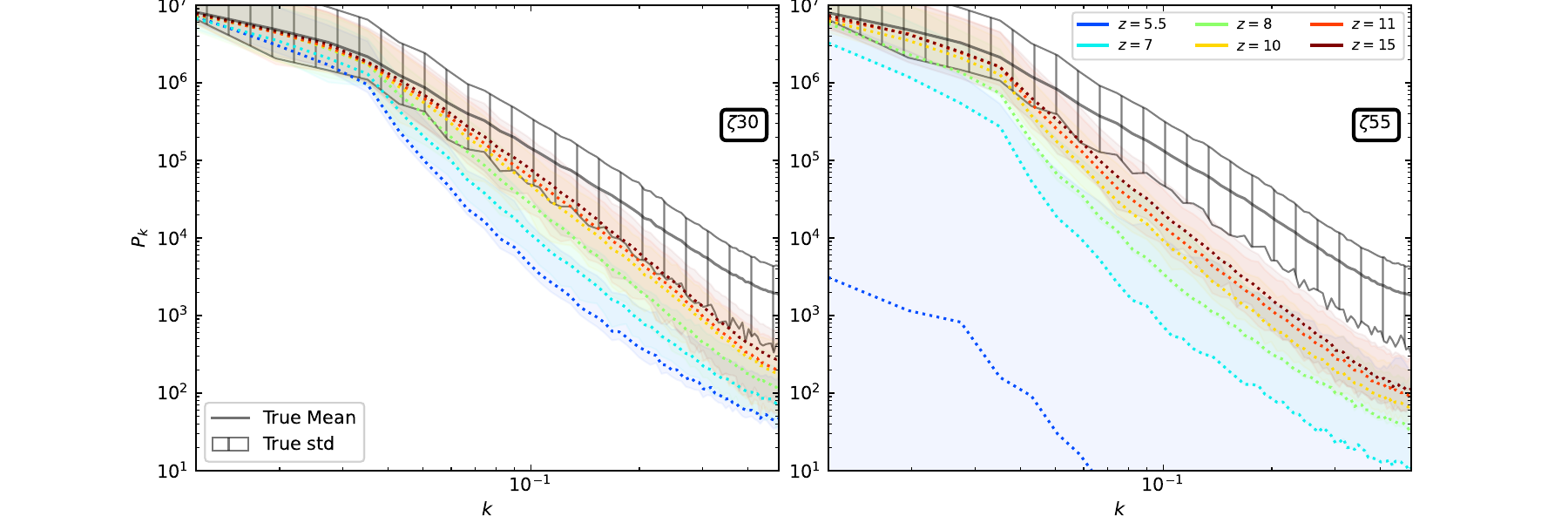}
    \caption{Power spectrum of the $\ZR$ field obtained with the model $\zeta 30$ on the left and $\zeta55$ on the right. k is the spatial frequency with dimension [cMpc$^{-1}h$]. The black line stands for the ground truth and the dotted coloured lines depict predictions with different training redshifts. The shaded areas depict the dispersion around the median (1${st}$ and 99${th}$ percentiles) for both prediction and ground truth.}
    \label{fig:Pkboth}
\end{figure*}

Eventually, Fig. \ref{fig:Pkboth} shows the power spectrum obtained with true and predicted $\ZR$ field. For $\zeta30$, the true power spectrum is closely followed, especially at large scale. However, the smaller scales turn off faster. This effect is even worse with $\zeta55$. For $\TR$ (Fig. \ref{fig:Pk30}), the smaller scales turn off more efficiently, 
%\LEt{***efficiently? "fair" suggests a subjective value judgement.      }
showing that small scales are more represented when working with cosmic time instead of redshift. 

The exact reason for this discrepancy is unclear. In \Cite{Thelie2022_2} we found that reionisation times are close to   Gaussian random fields (GRFs) and can be analysed by means of GRF theory, unlike $\ZR$ which is a non-linear function of $\TR$. We suspect that GRFs are more easily reconstructed as they provide, for instance, a symmetric distribution of values around the mean, whereas  $\ZR,$ for example, presents an asymetric distribution of values that is affected  the most from the smoothing of extrema inherent to our implementation of CNNs.

\end{appendix}
\end{document}